\documentclass[usenatbib]{mn2e}
\usepackage{mathtools}
\usepackage{graphicx}
\usepackage{epstopdf}
\usepackage{natbib}
\usepackage{verbatim}
\usepackage{amssymb,amsmath}
\usepackage{ulem}
\usepackage[T1]{fontenc}
\usepackage{aecompl}

\title[The Slow Flow Model]{The Slow Flow Model of Dust Efflux in Local Star-Forming Galaxies}

\author[Zahid et al.]{H.J. Zahid$^{1,2}$, P. Torrey$^{3}$, R.P. Kudritzki$^{1,4}$, L.J. Kewley$^{5}$, R. Dav\'e$^{6,7,8,9}$ \& M.J. Geller$^{2}$ \\ \\
$^{1}$University of Hawaii at Manoa, Institute for Astronomy - 2680 Woodlawn Drive, Honolulu,  HI 96822, USA \\
$^{2}$Smithsonian Astrophysical Observatory - 60 Garden St., Cambridge, MA 02138, USA \\
$^{3}$Harvard University, Astronomy Department - 60 Garden Street, Cambridge, MA 02138, USA \\
$^{4}$Max-Planck-Institute for Astrophysics - Karl-Schwarzschild-Str. 1, D-85741 Garching, Germany \\
$^{5}$Australian National University, Research School of Astronomy and Astrophysics - Cotter Road, Weston Creek, ACT 2611, Australia \\
$^{6}$ University of the Western Cape, Bellville - Modderdam Rd, Cape Town 7535, South Africa \\
$^{7}$ South African Astronomical Observatories - Observatory, Cape Town 7925, South Africa \\
$^{8}$ African Institute for Mathematical Sciences - Muizenberg - Atlantic Rd, Cape Town 7945, South Africa \\
$^{9}$University of Arizona, Department of Astronomy - 933 North Cherry Avenue, Rm. N204, Tucson, AZ 85721\\}

\begin{document}
\maketitle

\begin{abstract}

We develop a dust efflux model of radiation pressure acting on dust grains which successfully reproduces the relation between stellar mass, dust opacity and star formation rate observed in local star-forming galaxies. The dust content of local star-forming galaxies is set by the competition between the physical processes of dust production and dust loss in our model. The dust loss rate is proportional to the dust opacity and star formation rate. Observations of the relation between stellar mass and star formation rate at several epochs imply that the majority of local star-forming galaxies are best characterized as having continuous star formation histories. Dust loss is a consequence of sustained interaction of dust with the radiation field generated by continuous star formation. Dust efflux driven by radiation pressure rather than dust destruction offers a more consistent physical interpretation of the dust loss mechanism. By comparing our model results with the observed relation between stellar mass, dust extinction and star formation rate in local star-forming galaxies we are able to constrain the timescale and magnitude of dust loss. The timescale of dust loss is long and therefore dust is effluxed in a ``Slow Flow". Dust loss is modest in low mass galaxies but massive galaxies may lose up to $70\sim80\%$ of their dust over their lifetime. Our Slow Flow model shows that mass loss driven by dust opacity and star formation may be an important physical process for understanding normal star-forming galaxy evolution.

\end{abstract}


\section{Introduction}

\subsection{Background}


Dust is a fundamental constituent of the interstellar medium (ISM) of galaxies. Dust forms from material recycled back to the ISM through stellar winds or supernovae. The AGB phase of intermediate mass stars ($1\lesssim M_\odot \lesssim 8$) and massive stars ($\gtrsim 8 M_\odot$) which end their lives as Type II supernovae (SNe) are considered the dominant source of stellar dust production in star-forming galaxies while dust in the ISM may also form \textit{in situ} from accretion of enriched gas processed by stars \citep{Dwek1998}. Dust forms from metals and a strong correlation between gas-phase oxygen abundance and dust content is observed in local \citep{Heckman1998, Boissier2004, Asari2007, Garn2010b, Xiao2012, Zahid2012b} and high redshifts galaxies \citep{Reddy2010, Dominguez2013}.

Dust is composed of heavy elements and therefore the chemical and dust properties of galaxies should evolve consistently. The heavy element content of star-forming galaxies is characterized by a strong relation between the stellar mass and average gas-phase oxygen abundance \citep{Lequeux1979, Tremonti2004}. This so-called mass-metallicity relation (MZR) extends to low stellar mass galaxies \citep[$\sim10^7 M_\odot$][]{Lee2006, Zahid2012a, Berg2012} and is observed at intermediate \citep{Savaglio2005, Cowie2008, Zahid2011a, Moustakas2011, Zahid2013b} and high redshifts \citep{Erb2006b, Mannucci2009, Laskar2011}. The metallicity at all stellar masses increases with time and the high mass end of the relation flattens at late times as galaxies enrich to an empirical upper limit in the gas-phase abundance \citep{Zahid2013b}.

Examination of the second parameter dependencies of the MZR has revealed a correlation between stellar mass, metallicity and star formation rate \citep{Ellison2008, Lara-Lopez2010, Mannucci2010, Yates2012, Andrews2013}. We refer to this relation as the MZSR. For less massive galaxies, \citet{Lara-Lopez2010} and \citet{Mannucci2010} find that metallicity is \textit{anti-}correlated to SFR at a fixed stellar mass. However, the relation between stellar mass, metallicity and SFR is dependent on methodology \citep{Yates2012, Andrews2013}. Like \citet{Mannucci2010} and \citet{Lara-Lopez2010}, \citet{Yates2012} find a similar relation for low mass galaxies. However, they show that for massive galaxies the metallicity and SFR are \textit{positively} correlated at a fixed stellar mass. Thus, there is a ``twist" in the relation between stellar mass, metallicity and SFR. Not surprisingly, in these same galaxies, a similar twist is observed in the relation between stellar mass, dust extinction and SFR \citep{Zahid2013a}. Dust is formed from metals and we argue for a common physical origin for both relations. 

Recent observations indicate that stellar mass growth of most star-forming galaxies is dominated by secular evolution over the last $\sim$10 billion years. A strong correlation between stellar mass and SFR is observed out to at least $z\sim2.5$ \citep{Noeske2007a, Salim2007, Elbaz2007, Daddi2007, Pannella2009, Whitaker2012, Reddy2012} and is found to be independent of environment \citep{Peng2010, Wijesinghe2012, Koyama2013}. The small scatter ($\sim0.25$ dex) of the relation is independent of redshift suggesting that secular processes such as cosmological gas accretion dominate over mergers in building up the stellar mass of galaxies since at least $z\sim2-3$. The observed stellar mass-SFR relation (MSR) places strong constraints on the star formation histories of galaxies \citep{Noeske2007b, Conroy2009b, Leitner2012, Zahid2012b}. The observed MSR at several epochs implies that the majority of local star-forming galaxies are best characterized by continuous star formation histories. The observations demand that the majority of local star-forming galaxies maintain SFRs that do not vary significantly from their mean SFRs over most of their lifetime \citep{Noeske2007a}. 

The well defined MSRs and MZRs observed over cosmological time provide purely empirical constraints for continuous stellar mass growth and chemical enrichment of galaxies as they evolve \citep[see][]{Zahid2012b}. We attempt to understand trends in the metal and dust distribution of local galaxies in the context of this empirical framework. We develop a model where trends in the metal and dust distribution are naturally explained as a consequence of star-formation and large scale galactic gas flows. Recent theoretical considerations suggest that momentum, unlike energy which can be radiated away, may be the primary driver of large scale gas flows \citep{Murray2005}. Dust plays a crucial role since radiation couples to dust over a large and continuous range of wavelengths \citep{Draine2003} and provides a convenient mechanism for momentum transfer between the radiation field and the gas \citep[e.g.][]{Murray2005}.    


The potential importance of radiation pressure acting on dust in the context of galactic mass loss is well recognized. \citet{Chiao1972} suggest that under the influence of radiation pressure, dust may escape galaxies along magnetic field lines. \citet{Ferrara1993} posits that dusty diffuse clouds embedded in a anisotropic radiation field will feel a net acceleration due to radiation pressure. \citet{Davies1998} argue that driven by an imbalance between radiation and gravitational forces, disk galaxies may expel a large fraction ($90\%$) of dust produced over $\sim1$Gyr timescales. Dynamical coupling of dust and gas through collisions or coulomb interactions could provide a mechanism for transferring momentum from the radiation field to the gas \citep{Draine2004}. Using cosmological simulations, \citet{Aguirre2001} show that the IGM could be enriched by the expulsion of dust and gas driven by radiation pressure. Several models incorporating dust driven winds have recently been considered in the literature \citep{Zhang2010, Zu2011, Sharma2011, Sharma2012, Chattopadhyay2012, Wise2012}.

\subsection{A Model of Dust Driven Outflows}

We develop a numerical model of radiation pressure acting on dust grains leading to an expulsion of dust and metals as an explanation for the observed relation between stellar mass, dust extinction and SFR in SDSS galaxies shown in Figure \ref{fig:tau} \citep{Zahid2013a}. We refer to this relation as the MDSR. The relation between dust extinction and SFR changes with stellar mass. For galaxies at the same stellar mass dust extinction is \textit{anti-}correlated with the SFR at stellar masses $<10^{10} M_\odot$. There is a sharp transition in the relation at a stellar mass of $10^{10} M_\odot$. In massive galaxies dust extinction is \textit{positively} correlated with the SFR for galaxies at the same stellar mass.

The model we develop to reproduce the MDSR assumes the following:
\begin{enumerate}
\item At a fixed stellar mass the current SFR is \textit{anti-}correlated to the age of the galaxy.
\item The rate of dust production is proportional to the rate at which mass is recycled by stars.

\item The rate of dust loss is dependent on the amount of dust present and the rate at which high energy photons are produced.

\item The timescale of dust loss is comparable but not identical to the timescale of dust production.

\end{enumerate}
Under these assumptions we propose the following as an interesting physical model explaining the observed MDSR: Radiation pressure driven galactic mass loss \citep[e.g.][]{Murray2005} is a ubiquitous process in normal star-forming galaxies. Momentum is physically deposited into the ISM by the absorption of high energy photons. Therefore the galactic mass loss rate is proportional to the opacity and SFR (Assumption 3). The main source of opacity in star-forming galaxies is dust and in low mass galaxies dust can accumulate because the opacities, SFRs and therefore galactic mass loss rates are all low. The dust content of a galaxy is proportional to the total amount of stellar mass recycled (Assumption 2). Because stellar mass recycling is a time dependent process, the total amount of stellar mass recycled in a galaxy is correlated to galaxy age; older galaxies necessarily recycle a greater fraction of their stellar mass. At a fixed stellar mass the current SFR is \textit{anti}-correlated to galaxy age (Assumption 1). Therefore dust opacity is \textit{anti-}correlated with SFR for lower mass galaxies. As galaxies build up stellar mass they accumulate dust. This leads to an increase in their mass loss rate. If the timescale of dust loss is the same as dust production then galaxies should reach an equilibrium where dust loss is balanced by dust production. However, if the timescale over which dust is expelled from the ISM is not \textit{identical} to the timescale of dust production (Assumption 4), then galaxies with high stellar mass recycling rates (SMRRs), and therefore high dust production rates (Assumption 2), may accumulate dust. At a fixed stellar mass the SMRR is positively correlated with the SFR and therefore we expect a similar correlation between the dust content and SFR. A model such as this accounts for all the notable features in the observed MDSR. Because the timescale for dust loss is long, we refer to this as the ``Slow Flow" process.


We develop the Slow Flow model by bringing together star formation, chemical evolution and time dependent stellar mass recycling in order to understand the dust properties of local star-forming galaxies. In Section 2 we discuss the data and present the observed MDSR. In Section 3 and 4 we describe our numerical implementation of Assumptions 1 and 2, respectively. We describe our implementations of Assumptions 3 and 4 and demonstrate that we can reproduce the MDSR in Section 5. In Section 6 we justify our interpretation of dust loss as dust efflux in a slow flow rather than dust destruction. We revisit the assumptions in Section 7. In Section 8 we provide a discussion and we conclude with a summary of our model in Section 9. Table \ref{tab:key_acr} and \ref{tab:key_sym} provide a key for frequently used acronyms and symbols, respectively. Throughout this work we adopt the standard cosmology $(H_{0}, \Omega_{m}, \Omega_{\Lambda}) = (70$ km s$^{-1}$ Mpc$^{-1}$, 0.3, 0.7) and a \citet{Chabrier2003} initial mass function (IMF).

\begin{table}
\begin{center}
\caption{Key for Frequently Used Acronyms}
\label{tab:key_acr}
\begin{tabular}{c l}

Acronym & Definition \\

\hline
\hline

SFR & star formation rate \\
SFH & star formation history \\
ISM & interstellar medium \\
IMF & initial mass function \\
MZR & the stellar mass-metallicity relation \\
MSR & the stellar mass-star formation rate \\
        & relation  \\
SMRR & stellar mass recycling rate \\
MDSR & the stellar mass-dust extinction- \\
           & star formation rate relation \\
MZSR & the stellar mass-metallicity- \\
           & star formation rate relation \\
\hline
\hline
\end{tabular}
\end{center}
\end{table}

\begin{table}
\begin{center}
\caption{Key for Frequently Used Symbols}
\label{tab:key_sym}
\begin{tabular}{c l}
Symbol & Definition \\
\hline
\hline
$M_\ast$ & stellar mass \\
$\Psi$ & star formation rate \\
$\tau$ & dust opacity \\
$\dot{M}_R$ & mass recycling rate \\
$M_R$ & total mass of recycled gas \\
$t$ & time, variable of integration \\
$t_f$ & formation time of galaxy \\
$S$ & offset from the observed median relation \\
        & between stellar mass and star formation rate \\
$f_{mr}$ & fraction of mass recycled as a function of \\
               & time for single burst stellar population \\
$\dot{f}_{mr}$ & time derivative of fraction of mass recycled \\
$M_d$ & mass of dust \\ \\ \\

Free Parameters \\
$\Delta t$ & timescale of dust loss/destruction \\
$\eta$ & efficiency of dust loss in slow flow \\
$\alpha$ & efficiency of dust loss in fast flow \\
\hline
\hline
\end{tabular}
\end{center}
\end{table}

\section{Data and Methods }

\subsection{The MDSR}

In \citet{Zahid2013a} we present the observed MDSR for $\sim150,000$ star-forming galaxies in the SDSS DR7 \citep{Abazajian2009}. We summarize the data selection and methodology and refer the reader to \citet{Zahid2013a} for more details. We adopt the stellar masses and SFRs given in the DR7. The stellar masses are determined from the $ugriz$-band photometry \citep{Stoughton2002}. The SFRs are derived by the MPA/JHU group using the technique of \citet{Brinchmann2004} with additional improvements given by \citet{Salim2007}. The SFRs are corrected for dust and aperture effects. 

We distinguish star-forming galaxies from AGN using the [OIII]$\lambda5007$/H$\beta$ vs [NII]$\lambda6584$/H$\alpha$ diagram \citep{Kauffmann2003, Kewley2006}. In order to obtain a robust estimate of the Balmer decrement, we require that the signal-to-noise of the H$\alpha$ and H$\beta$ line be greater than 8. These selection criteria give us a sample of $\sim150,000$ star-forming galaxies. A detailed analysis of selection effects is presented in \citet{Zahid2013a}.

\begin{figure*}
\begin{center}
\includegraphics[width=1.8\columnwidth]{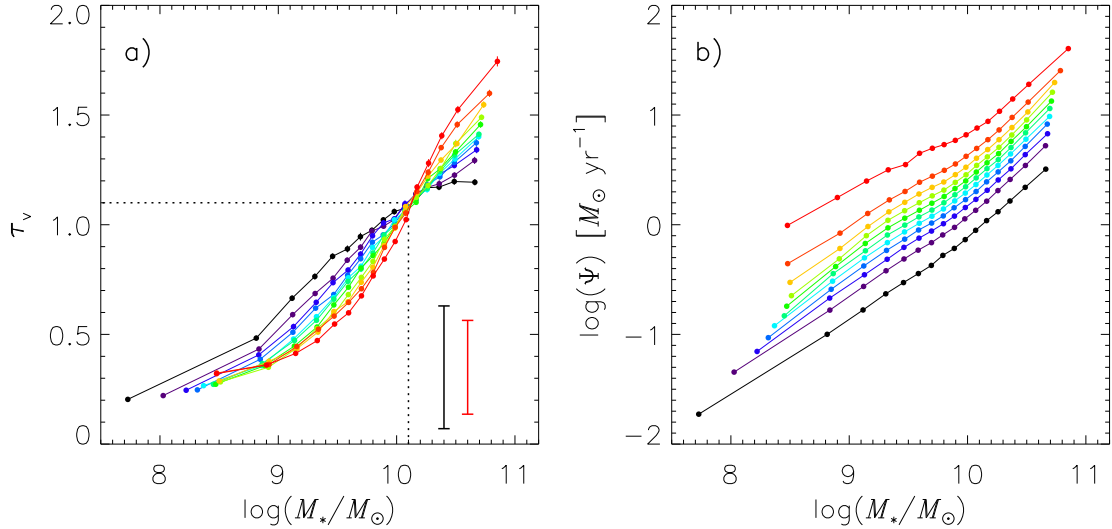}
\end{center}
\caption{The observed relation between stellar mass, dust opacity and SFR for a sample $\sim150,000$ galaxies from the SDSS DR7. a) The median optical depth derived from the Balmer decrement (see text for details) sorted into bins of stellar mass and SFR. The black and red error bars are the median scatter in each bin and median observational uncertainty, respectively. b) The corresponding SFR in each bin. Each curve is an undecile of the SFR in bins of stellar mass. The standard error for each bin is plotted but is typically smaller than the data point. \citep[This figure is a reproduction of Figure 2 from][]{Zahid2013a}}
\label{fig:tau}
\end{figure*}

Dust extinction is determined from the Balmer decrement assuming Case B recombination. For a gas with electron temperature T$_e$  = 10$^4$K and electron density $n_e = 10^2$ cm$^{-3}$, the intrinsic H$\alpha$/H$\beta$ ratio is expected to be 2.86 \citep{Osterbrock1989}. We derive the intrinsic colour excess, E(B$-$V), and the correction for dust attenuation using the extinction law of \citet{Cardelli1989} and a corresponding $R_\mathrm{v} = 3.1$. From the visual extinction, $A_\mathrm{v}$, we determine the visual optical depth from the relation $\tau_\mathrm{v} = A_\mathrm{v}/1.086 = R_\mathrm{v}$ E(B$-$V)/1.086. 

Figure \ref{fig:tau} \citep[c.f. Figure 2 from][]{Zahid2013a} shows the observed MDSR for local star-forming galaxies. The $\sim150,000$ SDSS galaxies in our sample are first sorted into 16 equally populated bins of stellar mass. Each of the 16 bins of stellar mass are then sorted into 11 equally populated bins of SFR. The data plotted in Figure \ref{fig:tau}a and \ref{fig:tau}b are the same (i.e. belong to the same bin of stellar mass and SFR). Figure \ref{fig:tau}a and \ref{fig:tau}b show the median optical depth and median SFR in bins of stellar mass and SFR. In each bin, the median is determined from $\sim890$ galaxies. The curves in Figure \ref{fig:tau}a are colour-coded to match the SFRs shown in Figure \ref{fig:tau}b. Each curve is an undecile\footnote{Each of eleven equal groups into which a population can be divided.} of the SFR in bins of stellar mass. The red curves correspond to the highest SFR in each bin of stellar mass and the black curves the lowest SFR bin in each bin of stellar mass. The median 1$\sigma$ scatter of the optical depth binned by stellar mass and SFR (Figure \ref{fig:tau}a) is 0.30 with 0.22 attributable to observational uncertainty. The error of the median optical depth in bins of stellar mass and SFR is determined from bootstrapping and is analogous to the standard error on the mean. The standard error for the optical depth in each bin of stellar mass and SFR is $\sim0.01$. The error for each bin is plotted in Figure \ref{fig:tau}a but is typically smaller than the data points.

As Figure \ref{fig:tau} shows, the dust opacity is correlated with stellar mass. Dust is formed from material recycled back into the ISM from stars. Massive galaxies have recycled a larger amount of material and therefore tend to be dustier. More notable is the relation between dust extinction and SFR which changes with stellar mass. For galaxies at the same stellar mass dust extinction is \textit{anti-}correlated with the SFR for galaxies with stellar masses $<10^{10} M_\odot$. At a stellar mass of $10^{10}M_\odot$ there is a sharp transition. In massive galaxies dust extinction is \textit{positively} correlated with the SFR for galaxies at the same stellar mass. \citet{Yates2012} see a similar ``twist" in the relation between metallicity and SFR for star-forming galaxies in the local universe \citep[for details see][]{Zahid2013a}.

\subsection{The Stellar Mass, Metallicity and Dust Extinction Relation}

Dust is formed from metals and therefore a correlation between metallicity and extinction is expected. \citet{Xiao2012} show that the colour excess is well fit as a function of stellar mass and metallicity. In \citet{Zahid2012b} we derive a fit to the colour excess which is
\begin{equation}
\mathrm{E(B-V)} = (0.12 + 0.041 Z^{0.77}) \times M^{0.24},
\label{eq:ebvfit}
\end{equation}
where $Z = 10^{(12 + \mathrm{log(O/H)} - 8)}$ and $M = M_{\ast}/(10^{10} M_\odot$). The metallicity is determined from the strong line method of \citet{Kobulnicky2004}. There is a well known discrepancy between various methods of determining gas-phase metallicities \citep[e.g.][]{Kennicutt2003, Kewley2008}. However, our analysis only requires \textit{relative} accuracy in estimating metallicity which the \citet{Kobulnicky2004} calibration delivers \citep[see][]{Kewley2008}. The fit is to $\sim20,000$ SDSS galaxies \cite[for more information on this sample see appendix of][]{Zahid2011a} and the RMS of the fit is 0.11 dex.


\subsection{Galaxy Age}

The spectrum of a galaxy contains a significant amount of information relating to its physical properties and the features of the galaxy spectrum are often interpreted using evolutionary stellar population synthesis models. The shape of the continuum within a galaxy is directly related to its underlying stellar population and the chemical properties of the stellar population can be inferred from the absorption line features. A standard set of absorption indices have been calibrated for this task \citep[e.g.][]{Worthey1994}. \citet{Tojeiro2007} derive a method, dubbed versatile spectral analysis (VESPA), to determine star formation and metallicity histories of galaxies from the shape of the continuum and standard absorption indices. A catalogue with VESPA applied to the SDSS DR7 galaxies is publicly available \citep{Tojeiro2009}\footnote{http://www-wfau.roe.ac.uk/vespa/}. In order to estimate the ages of galaxies we adopt the values determined from the stellar population synthesis models of \citet{Maraston2005} with a one-parameter dust model. In the VESPA catalogue, the star formation histories (SFHs) of galaxies are determined in 16 logarithmically spaced bins in lookback time. The total amount of stellar mass formed in each bin is given. We adopt the centre of each bin interval as the lookback time for each bin. The average age of stars for each galaxy is then determined from the mass-weighted lookback time.

\section{A Model for Stellar Mass Growth}

We assume that at a fixed stellar mass the current SFR of galaxies is \textit{anti-}correlated with the age of the stellar population (Assumption 1). Stochasticity in the star-formation process demands that this is not necessarily true for individual galaxies. However, each bin in Figure \ref{fig:tau} is the median of $\sim890$ galaxies. We emphasize that our model assumption \textit{only} applies to galaxies in an average sense; on average, at a fixed stellar mass, galaxies with higher SFRs have formed their stars over a shorter timescale and are necessarily younger.  

\begin{figure}
\begin{center}
\includegraphics[width=\columnwidth]{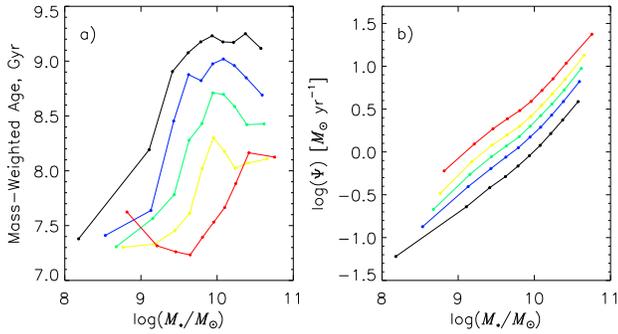}
\end{center}
\caption{a) The mass-weighted age determined from the VESPA models for $\sim135,000$ galaxies in the SDSS DR7 sorted into bins of stellar mass and SFR. b) The corresponding SFR for each bin of mass-weighted age.}
\label{fig:age}
\end{figure}

The observational data support the assumption that in the galaxy population-on average-the SFR is anti-correlated to the age. In Figure \ref{fig:age} we plot the relation between stellar mass, SFR and mass-weighted age. These data are a sub-sample of the $\sim150,000$ SDSS galaxies used to determine the MDSR for which we are able to measure mass-weighted ages using the VESPA models \citep{Tojeiro2009}. The sample is comprised of $\sim135,000$ galaxies. We use a binning procedure similar to the one used for deriving the MDSR. The data are first sorted into 10 equally populated bins of stellar mass and then each bin of stellar mass is sorted into 5 equally populated bins of mass-weighted age. Each bin contains $\sim2700$ galaxies. Figure \ref{fig:age} shows that massive galaxies have older stellar populations \citep[see also][]{Kauffmann2003a}. More importantly, at a fixed stellar mass the age of a galaxy is \textit{anti}-correlated to its SFR. 

We implement Assumption 1 numerically by requiring that galaxies evolve along the observed MSR \citep[among others]{Noeske2007a, Salim2007, Daddi2007, Elbaz2007, Pannella2009, Whitaker2012}. In order to produce an \textit{anti}-correlation between stellar mass and galaxy age, we populate the scatter in the MSR such that galaxies maintain a constant offset as they evolve. \citet{Whitaker2012} show that the intrinsic $1\sigma$ scatter in the MSR is $\sim0.25$ dex and the slope slightly flattens at higher redshift. There is significant evolution in the zero point. We adopt a constant slope for the MSR. Because redshift evolution of the MSR is dominated by evolution in the zero point, our adoption of a constant slope for the MSR does not affect our results.

\begin{figure*}
\begin{center}
\includegraphics[width=1.5\columnwidth]{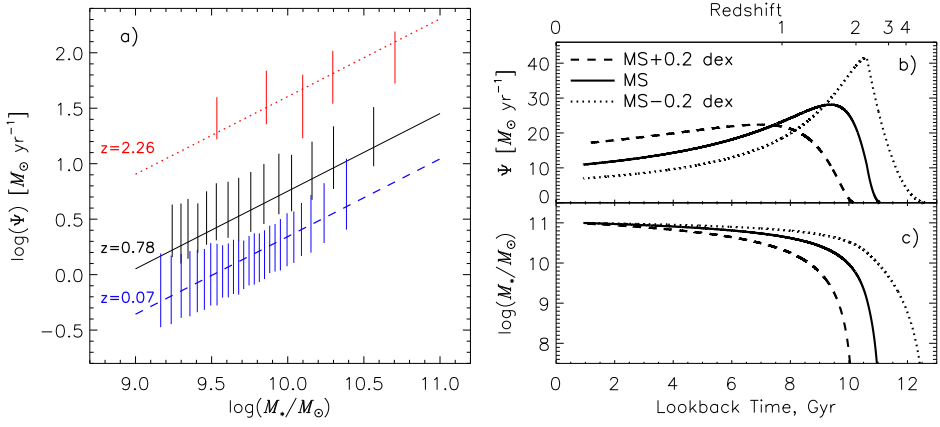}
\end{center}
\caption{a) The MSR at three epochs. The error bars plot the $1\sigma$ scatter of the SFR in bins of stellar mass. b) The star formation and c) stellar mass history for three model galaxies determined from integrating Equation \ref{eq:sfr_scatter} assuming the SFRs given by Equation \ref{eq:sfr}. Each galaxy has a stellar mass of $10^{11} M_\odot$ but with different values adopted for the offset from the MSR ($S$ in Equation \ref{eq:sfr_scatter}).}
\label{fig:sfr_mass}
\end{figure*}

In Figure \ref{fig:sfr_mass}a we plot the MSR at three redshifts. The data at $z = 0.07$ and 0.8 are taken from \citet{Zahid2011a} and the $z=2.26$ data come from \citet{Erb2006b}. These data are compiled in Table 1 of \citet{Zahid2012b}. From these data we derive the SFR as a function of stellar mass and redshift. In \citet{Zahid2012b} we show that the derived relation is consistent with other studies. The SFR as a function of stellar mass and redshift is
\begin{equation}
\Psi(M_\ast, z) = 2.00 \cdot \mathrm{exp}(1.33 z) \left( \frac{M_\ast}{10^{10}}\right)^{0.7} [M_\odot \, \mathrm{yr}^{-1}].
\label{eq:sfr}
\end{equation}
Observations of the MSR at higher redshifts suggest that the zero point does not evolve significantly beyond $z>2.3$ \citep[see][and references therein]{Dutton2010} and in our model calculations we assume no evolution in the zero point and stellar mass slope, i.e. $\Psi(M_\ast, z>2.26) = \Psi(M_\ast, 2.26)$. In our numerical implementation we transform $\Psi(M_\ast, z)$ to $\Psi(M_\ast, t)$ using the standard conversion between redshift and time implemented in the astronomy users library IDL routine \textit{galage.pro}.

The SFHs of galaxies can be derived from a simple model requiring that galaxies evolve along the mean MSR at all epochs \citep{Leitner2012, Zahid2012b}. In the simplest analytical model the rate of stellar mass growth is given by 
\begin{equation}
\frac{dM_\ast}{dt} = \Psi - \dot{M}_R,
\label{eq:mdot}
\end{equation}
where $\Psi$ is the SFR and $\dot{M}_R$ is the rate that mass is recycled to the ISM through various stellar mass loss processes. The stellar mass of a galaxy is then given by 
\begin{equation}
M_\ast(t) = \int_{t_f}^{t} \left[\Psi(t^\prime) - \dot{M}_R(t^\prime) \right]\, dt^\prime + M_{\ast,f}.
\label{eq:sfh}
\end{equation}
Here $M_{\ast,f}$ is the stellar mass at $t_f$. If $M_{\ast,f}$ is set to some arbitrarily low value (in our models $M_{\ast,f} = 10^6 M_\odot$) then $t_f$ can be interpreted as the formation time of the galaxy. In order to implement Assumption 1, we build on this model by requiring that a galaxy populate the scatter in the MSR such that it maintains a constant offset at all epochs. Analytically this is given by
\begin{equation}
M_\ast(t) = \int_{t_f}^{t} \left[S\Psi(t^\prime) - \dot{M}_R(t^\prime) \right]\, dt^\prime + M_{\ast,f},
\label{eq:sfr_scatter}
\end{equation}
where $S$ is a constant offset accounting for the scatter in the MSR. The formation time, $t_f$, and $S$ uniquely define the SFH of a galaxy in our model. 

In Figure \ref{fig:sfr_mass}b and \ref{fig:sfr_mass}c we show the star formation and stellar mass history, respectively, for three model galaxies that all have a stellar mass of $10^{11} M_\odot$ in the local universe ($t = 13$ Gyr in Equation \ref{eq:sfr_scatter}). The model galaxies in Figure \ref{fig:sfr_mass} evolve such that they have a constant offset relative to the MSR at all epochs (log($S$) = -0.2, 0 and 0.2 in Equation \ref{eq:sfr_scatter}). The stellar mass as a function of time for these model galaxies is determined from numerically integrating Equation \ref{eq:sfr_scatter}. 

A fundamental feature of this model of stellar mass growth is that higher mass galaxies are older \citep[c.f.][]{Noeske2007b}. Furthermore, at a fixed stellar mass, galaxies with higher SFRs are younger than galaxies with lower SFRs. This can be seen by the comparing the onset of star formation for the three model galaxies in Figure \ref{fig:sfr_mass}b. All three galaxies have a stellar mass of $10^{11}M_\odot$ but have different times for the initial onset of star formation, $t_f$. In our model the age of the galaxy ($t - t_f$) is \textit{anti}-correlated to the offset from the MSR, $S$. Consequently, at a fixed stellar mass the age of our model galaxies is \textit{anti-}correlated to the current SFR.

\section{A Model for Dust Formation}

\begin{figure}
\begin{center}
\includegraphics[width=\columnwidth]{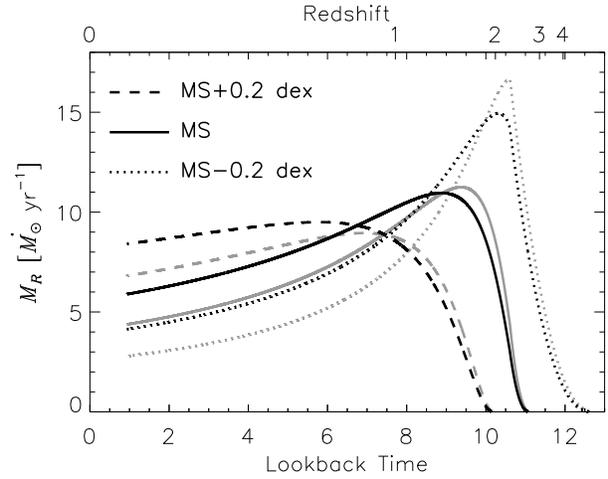}
\end{center}
\caption{The SMRR (black curves) as a function of time for the same three model galaxies in Figure \ref{fig:sfr_mass}. The SFRs are given by the grey curves for reference and have been scaled by a factor of 0.4 for ease of comparison.}
\label{fig:mrr}
\end{figure}

Assumption 2 of our model is that the rate of dust production is proportional to the stellar mass recycling rate. A large amount of dust in star-forming galaxies is thought to be produced by massive ($\gtrsim 8 M_\odot$) stars that end their lives as supernovae and post main-sequence evolution (AGB phase) of intermediate mass ($1 \lesssim M_\odot \lesssim 8$) stars \citep{Dwek1998}. Because all dust forms from heavier elements, dust that may form in the ISM is also dependent on the gas processed and recycled by stars. Dust production is therefore intimately related to the physical process of stellar mass recycling. In our numerical implementation, we adopt the simplest assumption that the dust production rate is directly proportional to the SMRR.

We implement continuous, time dependent stellar mass recycling following \citet{Jungwiert2001}. They give the SMRR as
\begin{equation}
\dot{M}_{R}(t) = \int_0^t \Psi(t^\prime) \dot{f}_{mr}(t - t^\prime) dt^\prime. 
\label{eq:mrr}
\end{equation}
Here $\dot{M}_R(t)$ is the SMRR as a function of time, $\Psi(t)$ is the SFR and $\dot{f}_{mr}(t)$ is the time derivative of the fraction of mass recycled to the ISM at time $t$ for a single, instantaneous burst stellar population. The rate at which stellar mass is recycled to the ISM is given as a convolution of the SFR with the time derivative of the fractional mass recycled. \citet{Jungwiert2001} parameterize the fractional mass recycled from a single, instantaneous burst stellar population as a function of time by
\begin{equation}
f_{mr}(t) = C_0 \, \mathrm{ln} \left(\frac{t}{\lambda} + 1 \right).
\label{eq:return}
\end{equation}
Both $C_0$ and $\lambda$ are constants depending on the particular choice of IMF \citep[values are given in Table 1 of][]{Leitner2011}. For a Chabrier IMF $C_0 = 0.046$ and $\lambda = 2.76 \times 10^5$ [yrs]. Because there is an interdependency between the SMRR and SFH, we iteratively determine both quantities using Equation \ref{eq:sfr_scatter} \citep{Leitner2011, Zahid2012b}. In Figure \ref{fig:mrr} we plot the SMRR for the three model galaxies shown in Figure \ref{fig:sfr_mass}b and \ref{fig:sfr_mass}c. Note that due to the convolution, the SMRR is offset in time from the SFR.

\section{Results: An Analytical Model Reproducing the Observed Relation Between Stellar Mass, Dust Extinction and SFR}

\subsection{Dust Model Outputs}

\begin{figure*}
\begin{center}
\includegraphics[width=1.8\columnwidth]{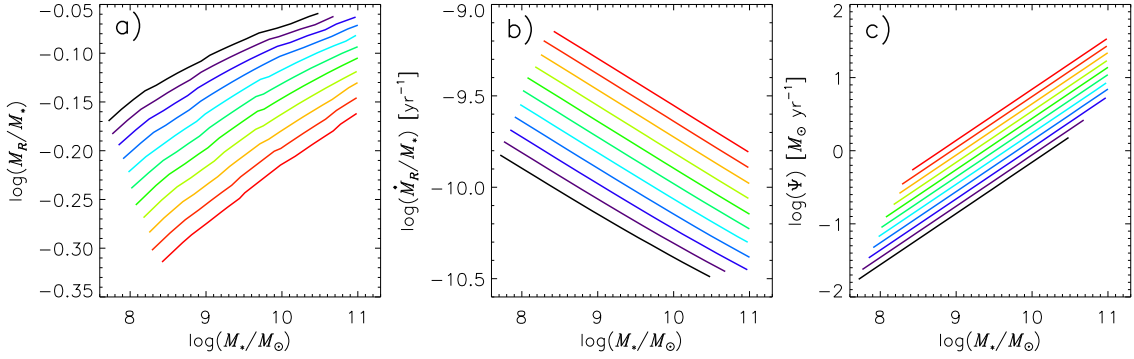}
\end{center}
\caption{The stellar mass recycling properties of model galaxies plotted against stellar mass. a) The total amount of stellar mass recycled divided by the current stellar mass. b) The specific SMRR (the SMRR divided by current stellar mass) and c) the corresponding SFR for each model galaxy. }
\label{fig:return}
\end{figure*}

We run our models adopting a range of values for the constant offset from the MSR, $S$, and formation time, $t_f$. The 2$\sigma$ scatter observed in the MSR is $\sim0.5$ dex. We run our models for 11 values of $S$ equally spaced such that $-0.5 \leq \mathrm{log}(S) \leq 0.5$. For the formation times we adopt a range of values corresponding to $z < 12$. We use only models with final stellar masses in the range of $8 \lesssim M_\ast/M_\odot \lesssim11$. This allows us to explore practically\footnote{In our models, the maximum mass a galaxy with -0.5 dex offset can form since $z=12$ is $\sim10^{10.5}M_\odot$ (see Figure \ref{fig:return}).} the full range of SFRs and stellar masses observed in our local sample (see right panel of Figure \ref{fig:tau}b). The following analysis is based on 220 model galaxies.

The total amount of stellar mass recycled to the ISM as a function of time is determined by integrating the SMRR over time which is
\begin{equation}
M_R(t) = \int_{t_f}^{t} \dot{M}_R(t^\prime) dt^\prime.
\label{eq:mr}
\end{equation}
In Figure \ref{fig:return}a we plot the total amount of stellar mass recycled (normalized to stellar mass) as a function of stellar mass. In Figure \ref{fig:return}b we show the specific SMRR which is the SMRR divided by stellar mass (analogous to specific SFR) as a function of stellar mass. In Figure \ref{fig:return}c we show the SFR as a function of stellar mass. In the figure each colour corresponds to a different offset from the MSR. The total amount of mass recycled to the ISM (Figure \ref{fig:return}a) is \textit{anti}-correlated to the SFR. This is a consequence of the fact that at a fixed stellar mass, galaxies with lower SFRs are older and therefore their stellar populations have had more time to recycle mass back to the ISM through stellar winds and supernovae. Conversely, the SMRR (Figure \ref{fig:return}b) is \textit{positively} correlated to the SFR. Because the SMRR is related to the SFR through the convolution given in Equation \ref{eq:mrr}, at a fixed stellar mass galaxies with higher SFRs also have higher SMRRs.

The \textit{anti}-correlation at a fixed stellar mass between the SFR and total amount of mass recycled and the \textit{positive} correlation between the SMRR and SFR seen in Figure \ref{fig:return} is the foundation for the physical model described in Section 1.2. We develop an analytical model such that at low stellar masses ($\lesssim10^{10} M_\odot$) the MDSR is governed by the total amount of stellar mass recycled and at higher stellar masses it is related to the current SMRR. Such a model naturally leads to an \textit{anti-}correlation between dust opacity and SFR at lower stellar masses and a \textit{positive} correlation between dust opacity and SFR at higher stellar masses.

\subsection{The Analytical Slow Flow Model}

In order to reproduce the observed MDSR, we develop an analytical model where the observed dust content of the galaxy is a competition between dust production and dust destruction/efflux. In the discussion that follows we refer to dust efflux and/or destruction as dust loss. The dust production rate is directly proportional to the SMRR (Assumption 2) and the rate of dust loss is proportional to the opacity and the SFR (Assumption 3). The timescale over which dust is lost is not necessarily coincidental with the dust production timescale (Assumption 4). Thus, we introduce a free parameter to account for a temporal offset between production and loss. The ``Slow Flow'' model is analytically given by
\begin{equation}
M_d(t) \propto \int_{t_f}^{t} \left[ \dot{M}_R(t^\prime) - \eta \, \tau(t^\prime) \, \Psi(t^\prime-\Delta t)\right] dt^\prime.
\label{eq:model}
\end{equation}
Equation \ref{eq:model} is the dust balance equation. $M_d(t)$ is the total amount of dust in the galaxy as a function of time. The first term on the right hand side is the dust production term which we assume is proportional to the SMRR, $\dot{M}_R$. The second term on the right hand side represents the rate at which dust is lost; $\tau$ is the opacity and $\Psi$ is the SFR. If $t^\prime - \Delta t < t_f$ then $\Psi(t^\prime - \Delta t) = 0$. $\dot{M}_R$ and $\Psi$ are determined from our model of stellar mass growth (Section 3) and dust production (Section 4). The free parameters of the model are $\eta$ which is the dust loss efficiency (the dust loss rate per unit SFR) and $\Delta t$, the timescale for dust loss.

\begin{figure*}
\begin{center}
\includegraphics[width=1.5\columnwidth]{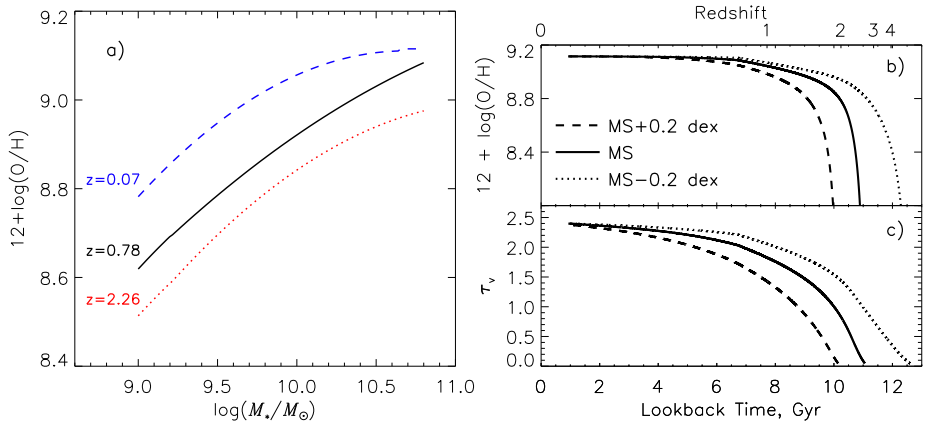}
\end{center}
\caption{a) The MZR at three redshifts \citep{Zahid2012b}. b) The metallicity and c) opacity as a function of time for the same three model galaxies as in Figure \ref{fig:sfr_mass}. The metallicites are determined by interpolating between the three MZRs and the opacity is determined from the relation given by Equation \ref{eq:ebvfit}. }
\label{fig:z}
\end{figure*}

In Equation \ref{eq:model}, we do not know $\tau$ a priori. For each model galaxy we derive $\tau$ as a function of time using the empirical relation given by Equation \ref{eq:ebvfit} and the observed redshift evolution of the MZR. In the left panel of Figure \ref{fig:z} we show the MZR at $z = 0.07, 0.8$ and 2.26 \citep{Zahid2012b}. We have used the \citet{Kobulnicky2004} calibration in determining metallicity though we emphasize that our results are independent of our choice of calibration. We determine the metallicity of a galaxy as a function of time by first determining its stellar mass history (see Section 3.1 and Figure \ref{fig:sfr_mass}c) and then using this to linearly interpolate the observed MZRs in time and stellar mass. This is shown for our three model galaxies in Figure \ref{fig:z}b. Using the empirical relation between extinction, stellar mass and metallicity given in Equation \ref{eq:ebvfit} we derive an optical depth as a function of stellar mass and time for the same model galaxies. This is shown in Figure \ref{fig:z}c.

In Figure \ref{fig:tau} we plot the MDSR. The dust content is given by the observed optical depth. Optical depth is given by $\tau = n \sigma_d L$ where $n$ is the volume density of particles, $\sigma_d$ is the geometric cross section for dust and $L$ is the line-of-sight path length. Our analytical model given in Equation \ref{eq:model} is proportional to the total mass of dust in the galaxy. In order to compare with observations we rescale our model such that 
\begin{equation}
\tau_{model} = A/M_\ast^{\beta} \int_{t_f}^{t} \left[ \dot{M}_R(t^\prime) - \eta \, \tau_\mathrm{v}(t^\prime) \, \Psi(t^\prime-\Delta t)\right] dt.
\label{eq:scale_model}
\end{equation}
Here a pre-factor $A/M_\ast^{\beta}$ is introduced to Equation \ref{eq:model}. We have the mass of dust $M_d \propto N$ where $N$ is the total amount of dust particles. To first order $N\sim nL^3$ where $L^3$ represents a volume. If we assume that $M_\ast \propto L^3$ then we have that $\tau \sim nL \sim N/M_\ast^{2/3} \sim M_d/M_\ast^{2/3}$. We set $\beta$ to 2/3. $A$ is an overall normalization accounting for geometric properties of the dust particles and is determined by normalizing the maximum value of $\tau$ in the models to the maximum observed optical depth in Figure \ref{fig:tau} ($\tau_{max} = 1.9$). The analytical model has four degrees of freedom: two constants to make our model output comparable to observations and two free parameters.

\subsection{The Parameter Space}
\label{ssec:pspace}

\begin{figure}
\begin{center}
\includegraphics[width=\columnwidth]{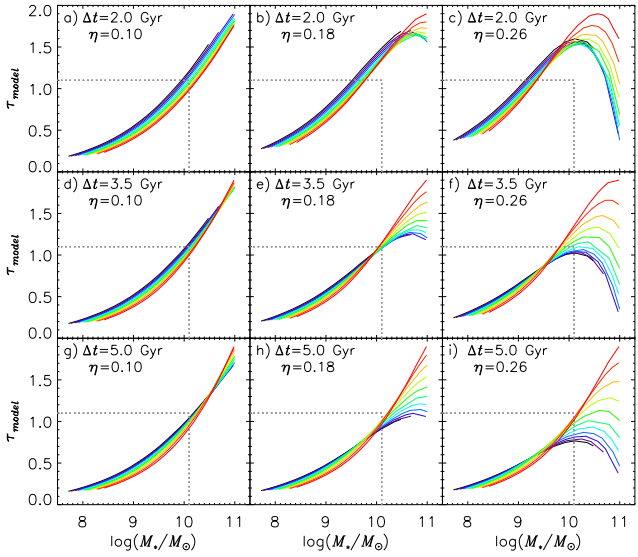}
\end{center}
\caption{The model MDSR as a function of the free parameters $\eta$ and $\Delta t$ in Equation \ref{eq:scale_model}. The colours correspond to the SFRs shown in Figure \ref{fig:return}c. The dotted lines indicate the stellar mass and opacity at which the twist in the observed MDSR (Figure \ref{fig:tau}) occurs.}
\label{fig:param}
\end{figure}

In Figure \ref{fig:param} we plot the MDSR determined from our model given in Equation \ref{eq:scale_model} for various choices of the free parameters. We adopt $\eta = [0.10, 0.15, 0.20]$ and $\Delta t = [2.0, 3.5, 5.0]$ Gyr and plot the models in the $3\times3$ panels of the figure. In each panel the curves plot $\tau$ as a function of stellar mass. $\tau$ is determined for each model galaxy by integrating from $t_f$ to $t=13$ Gyr in Equation \ref{eq:scale_model}. The curves are colour-coded to match the SFRs shown in Figure \ref{fig:return}c (e.g. the red curves are the high SFR galaxies and black curves are the low SFR galaxies). Our goal is to reproduce the observed MDSR shown in Figure \ref{fig:tau}.

In Figure \ref{fig:param}a we see that if $\eta \lesssim 0.1$ and $\Delta t \lesssim 2$ Gyr no twist is observed in the model MDSR. At a fixed stellar mass, galaxies with high SFRs are younger. Stellar mass recycling is a time dependent process and therefore younger galaxies have recycled a smaller fraction of their stellar mass back to the ISM. Because we take dust production to be proportional to the amount of recycled stellar mass, younger galaxies have less dust. When $\eta$ is small, dust loss is negligible and when $\Delta t$ is small, the relative amount of dust loss in low and high SFR galaxies, at a fixed stellar mass, is comparable. In this case there is no differential loss of dust in galaxies at a fixed stellar mass. Thus, in Figure \ref{fig:param}a galaxies that have high SFRs (e.g. red curve) have less dust than galaxies with low SFRs (e.g. black curve) because they have recycled less mass and therefore produced less dust.

The magnitude of the rate of dust loss is set by $\eta$. Increasing $\eta$ decreases the stellar mass at which the twist occurs in the model MDSR. Increasing $\eta$ also increases the spread of optical depths in galaxies with higher stellar masses independent of the value of $\Delta t$ adopted. This is seen when comparing rows of Figure \ref{fig:param} ($\eta$ increases from left to right). The dust loss term is (to first order) proportional to $\eta \tau \Psi$. The product of $\eta \tau$ sets the stellar mass at which dust loss becomes significant relative to dust production. Less massive galaxies have smaller $\tau$ and therefore increasing $\eta$ moves the twist to smaller stellar masses. Increasing the value of $\eta$ also leads to a larger spread in the optical depth at higher stellar masses because $\eta$, together with $\Delta t$, sets the relative amount of dust lost for galaxies at the same stellar mass. 

The maximum value of $\eta$ is limited by the fact that galaxies cannot lose more dust than they produce. This leads to the requirement that $\eta \tau \Psi \lesssim \dot{M}_R$ or $\eta \tau \lesssim \dot{M}_R/\Psi \sim 0.4$ for a Chabrier IMF. Panels c), f) and i) in Figure \ref{fig:param} show that for $\eta = 0.26$ high mass galaxies lose too much dust when compared to the observations in Figure \ref{fig:tau}. Thus the observations constrain $\eta$ such that $0.1 < \eta < 0.26$. 

The timescale of dust loss is set by $\Delta t$ and if it is shorter than the timescale of dust production/accumulation then no twist will be observed. For a twist to occur in the model MDSR $\Delta t > 2$ Gyr. Increasing $\Delta t$ from 2 to 3.5 Gyr decreases the stellar mass at which the twist occurs and slightly increases the spread in the optical depth at higher stellar masses. The timescale of dust loss, $\Delta t$, is the same for all galaxies but the timescale of dust production varies according to SFH. At a fixed stellar mass galaxies with higher SFRs have produced a greater fraction of their stars and dust more recently in cosmic time because they are younger (see Figure \ref{fig:age}). Thus there is a differential accumulation of dust depending on SFH. The differential accumulation of dust occurs on shorter timescales for higher mass galaxies because higher mass galaxies have higher SMRRs. Therefore, increasing $\Delta t$ moves the twist to smaller stellar masses and slightly increases the spread in optical depth. Increasing $\Delta t$ from 3 to 5 Gyr does not significantly alter the MDSR because galaxies typically reach a steady state in their SMRR within a few Gyr (see Figure \ref{fig:mrr}). 

In order to determine the best model parameters we attempt to qualitatively reproduce the observed MDSR shown in Figure \ref{fig:tau}. In particular, we choose parameters that best reproduce the location of the twist and the scatter in dust opacity for massive galaxies. A model with $\eta = 0.17$ and $\Delta t = 3.5$ Gyr best reproduces the observed relation. By varying the parameters we control the stellar mass at which the twist occurs and the relative spread in optical depths observed in high mass galaxies. The spread in the optical depth at low stellar masses (i.e. stellar masses less than the stellar mass where the twist occurs) is independent of our choice of parameters. Because $\tau$ is small for lower mass galaxies dust loss is inefficient and the spread in optical depth is set by the difference in age of galaxies at a fixed stellar mass. The model MDSR has a smaller spread in the values of $\tau$ at stellar masses $\lesssim10^{10} M_\odot$ when compared to the observed MDSR. Varying the IMF (e.g., using Chabrier Steep instead of Chabrier) or adding an additional dust loss term proportional the SFR (see below) can ease some of this tension between the model and observations at lower stellar masses.

The amount of dust produced is dependent on our choice of IMF in our model. Therefore, we may expect that the free parameters are as well. We perform a similar analysis using the Chabrier steep IMF which has a steeper high mass slope leading to $\sim20\%$ less mass loss \citep[see Figure 1 of][]{Leitner2011}. We are able to reproduce the observed MDSR using the same analytical model. The best value of $\eta$ is slightly smaller. $\Delta t$ is independent of our choice of IMF.

\subsection{Additional Parameters}
\label{ssec:param3}

\begin{figure}
\begin{center}
\includegraphics[width=\columnwidth]{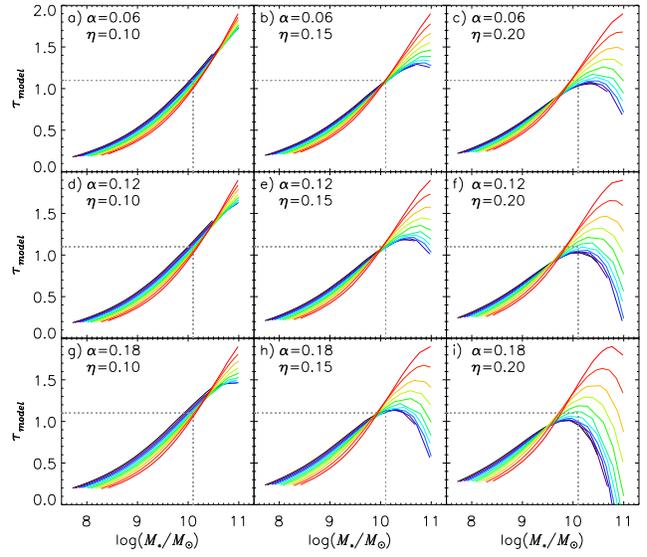}
\end{center}
\caption{The model MDSR as a function of the free parameters $\eta$ and $\alpha$ in Equation \ref{eq:model_alpha}. The colours correspond to the SFRs shown in Figure \ref{fig:return}c. The dotted lines indicate the stellar mass and opacity at which the twist in the observed MDSR (Figure \ref{fig:tau}) occurs.}
\label{fig:param3}
\end{figure}

We include an additional dust loss term to our model that is directly proportional the SFR. The analytical model presented in the previous section accounts for the notable features of the MDSR (Figure \ref{fig:tau}). However, additional physical processes temporally coincidental with the SFR may also be operating in galaxies. It is beyond the scope of this work to produce a model accounting for all possible physical mechanisms affecting the dust content of star-forming galaxies since the physics of dust production and evolution is still not well understood. The additional term added to the model is physically motivated and is meant to indicate the flexibility of the model.

\begin{figure*}
\begin{center}
\includegraphics[width=1.8\columnwidth]{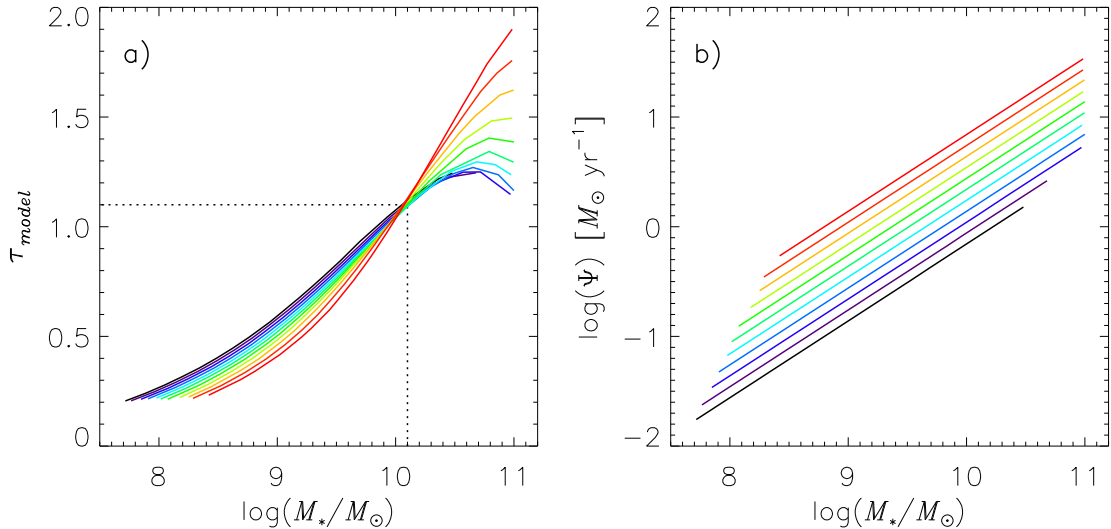}
\end{center}
\caption{a) The ``best-fit" model MDSR for the model given by Equation \ref{eq:model_alpha} with $\eta = 0.14$, $\alpha = 0.12$ and $\Delta t = 3.7$ Gyr. b) The SFRs corresponding to optical depths in shown in a).}
\label{fig:model_tau}
\end{figure*}

We modify the analytical model for dust production given in Equation \ref{eq:model} to include an additional term for dust loss that scales only with the SFR with no time delay. This is given by
\begin{equation}
M_d \propto \int_{t_f}^{t}  \left[ \dot{M}_R(t^\prime) -  \eta \, \tau(t^\prime) \, \Psi(t^\prime-\Delta t) - \alpha \Psi(t^\prime) \right] dt^\prime,
\label{eq:model_alpha}
\end{equation}
where we have introduced a new free parameter $\alpha$ which is a dust loss efficiency factor analogous to $\eta$ but for a physical mechanism that operates instantaneously and independent of dust optical depth. The term $\alpha \Psi$ could be interpreted physically as either dust loss or dust destruction. Thermal energy injected from Type II supernovae could drive outflows of dust in bubbles and along filaments of hot gas and/or destroy dust through UV photodissociation, sputtering and/or shocks. Following Equation \ref{eq:scale_model}, we scale the model given in Equation \ref{eq:model_alpha} by $A/M_\ast^{2/3}$.

In Figure \ref{fig:param3} we plot the MDSR determined from the three parameter model for various choices of the free parameters. Similar to the two parameter model, we find that when $\Delta t \gtrsim 3.5$ Gyr, the model MDSR is insensitive to $\Delta t$. We fix $\Delta t = 3.7$ Gyr and we adopt $\eta = [0.10, 0.15, 0.20]$ and $\alpha = [0.05, 0.10, 0.15]$. We plot models with these parameters in the $3\times3$ panels of the figure. In each panel the curves plot $\tau$ as a function of stellar mass. As before, the curves are colour-coded to match the SFRs shown in Figure \ref{fig:return}c (e.g. the red curves are the high SFR galaxies and black curves are the low SFR galaxies).

The new model given by Equation \ref{eq:model_alpha} gives a MDSR that is qualitatively similar to the model given by Equation \ref{eq:model}. Increasing $\eta$ decreases the stellar mass at which the twist occurs independent of $\alpha$. The effect of $\alpha$ is to increase the spread in optical depth at a fixed stellar mass at \textit{all} stellar masses. In producing the model MDSR, we normalize the relation such that the maximum observed optical depth is 1.9. Increasing the total amount of dust loss across all galaxies also increases the relative amount of dust loss for galaxies at a fixed stellar mass. The increased spread in optical depth in the model at all stellar masses eases some of the tension between the model and observations at low stellar masses.


Galaxies cannot lose more dust than they produce thus, to first-order, $\alpha \Psi + \eta \tau \Psi \lesssim \dot{M}_R$ or $\alpha + \eta \tau \lesssim \dot{M}_R/\Psi \sim 0.4$. In panel f) and i) the value of $\alpha + \eta \tau$ is too large and some galaxies lose more dust than they produce leading to an unphysical negative optical depth. A model with $\eta = 0.14, \alpha = 0.12$ and $\Delta t = 3.7$ Gyr best reproduces the observed MDSR. The primary result of this paper is shown in Figure \ref{fig:model_tau} and can be compared directly to the observed relation shown in Figure \ref{fig:tau}. We note that the model shown in Figure \ref{fig:model_tau} is very similar to the best model from Section \ref{ssec:pspace} (i.e. $\eta = 0.17$, $\Delta t = 3.5$ Gyr and $\alpha = 0$).

\subsection{Magnitude of Dust Loss}

\begin{figure}
\begin{center}
\includegraphics[width=\columnwidth]{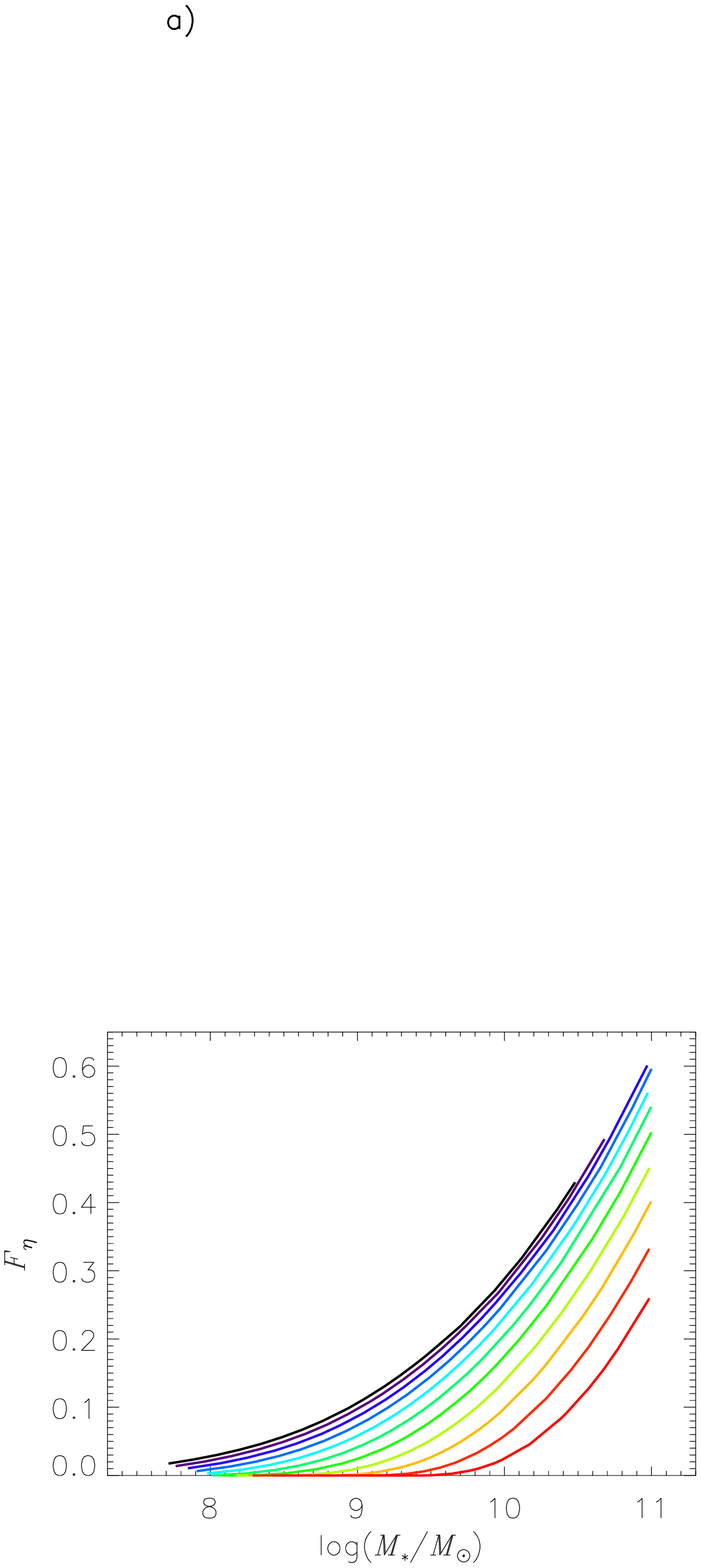}
\end{center}
\caption{The fraction of dust produced that is lost in the Slow Flow for the model given by Equation \ref{eq:model}. Fractional loss is calculated using Equation \ref{eq:slow}. The colours correspond to the SFRs shown in Figure \ref{fig:return}c.}
\label{fig:p2c}
\end{figure}

We are unable to estimate dust masses from our model since we derive the optical depth and not dust mass. However, we can estimate the fraction of dust lost (i.e., dust effluxed and/or destroyed) to the amount of dust produced in the model. For the two parameter model (Section 5.2) the fraction is given by
\begin{equation}
F_\eta = \frac{\int_{t_f}^{t} \eta \, \tau(t^\prime) \, \Psi(t^\prime-\Delta t) \, dt^\prime}{\int_{t_f}^{t}  \dot{M}_R(t^\prime) \, dt^\prime}.
\label{eq:slow}
\end{equation}
The numerator is proportional to the amount of dust lost and the denominator is proportional to the total amount of dust produced. We refer to quantity given by Equation \ref{eq:slow} as the slow loss term. In Figure \ref{fig:p2c} we plot $F_\eta$ as a function of stellar mass for the two parameter model. As before, the curves are colour-coded to match the SFRs shown in Figure \ref{fig:return}c. The fraction of dust lost, $F_\eta$, increases with stellar mass because dust loss is dependent on dust opacity and more massive galaxies are dustier. In this case, the magnitude of dust loss in low-mass galaxies is negligible; a property that is consistent with the galaxy formation model of \citep{Wise2012}. At a fixed stellar mass, galaxies with high SFRs lose a smaller fraction of their dust; they are younger and therefore have had less time to lose dust.

\begin{figure*}
\begin{center}
\includegraphics[width=2\columnwidth]{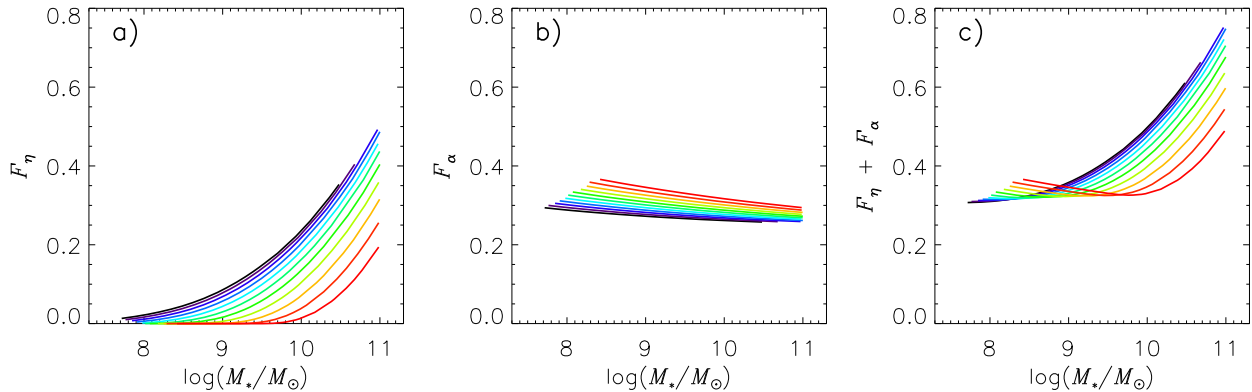}
\end{center}
\caption{Dust loss for the three parameter model given by Equation \ref{eq:model_alpha}. Fractional loss is calculated using a) Equation \ref{eq:slow} and b) Equation \ref{eq:fast}. c) The sum of the fractional loss given in a) and b). The colours correspond to the SFRs shown in Figure \ref{fig:return}c.}
\label{fig:p3c}
\end{figure*}

We also calculate the fraction of dust lost to dust produced for the three parameter model using Equation \ref{eq:slow} and
\begin{equation}
F_\alpha = \frac{\int_{t_f}^{t} \alpha \, \Psi(t^\prime) \, dt^\prime}{\int_{t_f}^{t}  \dot{M}_R(t^\prime) \, dt^\prime}.
\label{eq:fast}
\end{equation}
The numerator is proportional to the amount of dust lost and the denominator is proportional to the total amount of dust produced. We refer to the quantity given by Equation \ref{eq:fast} as the fast loss term. We plot $F_\eta$, $F_\alpha$ and $F_\eta + F_\alpha$ for our three parameter model discussed in Section \ref{ssec:param3} in Figure \ref{fig:p3c}a, b and c, respectively. The fraction of dust lost in the slow loss, $F_\eta$, for the three parameter model is $\sim10\%$ less than for our two parameter model because $\eta$ is $\sim 10\%$ smaller for the three parameter model. The amount of dust lost in the fast loss component is nearly constant with stellar mass because both the dust loss (numerator of Equation \ref{eq:fast}) and dust production (denominator of Equation \ref{eq:fast}) are proportional to the SFR. At a fixed stellar mass galaxies with lower SFRs are older and therefore have recycled a larger amount of their stellar mass back to the ISM. Thus, at a fixed stellar mass, galaxies with low SFRs have formed a greater amount of stars as compared to galaxies with high SFRs since galaxies with high SFRs have recycled a relatively smaller fraction of the material back to the ISM. The amount of dust produced is proportional to the amount of stellar mass recycled. Therefore, at a fixed stellar mass the \textit{anti-}correlation between $F_\alpha$ and SFR is due to the fact that younger galaxies (those with high SFRs) have produced less dust (see Figure \ref{fig:return}a).

\section{Dust Efflux as the Physical Basis of the Slow Flow Model}

Thus far, dust loss in the Slow Flow model has not been physically interpreted. In particular, the dust loss term parameterized by Equation \ref{eq:scale_model} could physically represent dust destruction and/or dust efflux. However, several lines of reasoning support an interpretation of dust efflux over dust destruction as the physical basis of the Slow Flow model.

The observed MZSR provides important constraints for physically interpreting the Slow Flow model developed to reproduce the MDSR. While dust may be destroyed, metals can not and therefore the observed MDSR is unlikely to be explained by destruction processes unless the similar trends observed in the MDSR and MZSR are taken to be coincidental \citep[see][]{Zahid2013a}. Furthermore, if dust destruction were responsible for the MDSR we may expect an opposite trend in the MZSR since destruction of dust particles would liberate the constituent metals, thus increasing the gas-phase abundance while decreasing the dust opacity. In light of similar trends in the MZSR derived by \citet{Yates2012}, the twist observed in the MDSR may be more naturally explained by dust efflux rather than dust destruction.

It is important to note that the MZSR is dependent on the methodology applied in deriving the physical properties of galaxies \citep[c.f.][]{Mannucci2010, Yates2012, Andrews2013}. In particular, there is a well-known and long-standing discrepancy in metallicities derived using various strong-line calibrations \citep[e.g.][]{Kennicutt2003, Bresolin2004, Nagao2006, Kewley2008, Moustakas2010}. For the same galaxies, theoretical calibrations typically yield abundances that are $\gtrsim0.3$ dex larger than empirical calibrations \citep{Kewley2008}. Much of the discrepancy may be attributed to the method used in calibrating the strong-line ratios. Empirical methods calibrate strong-line ratios against the metallicities derived from temperature sensitive auroral lines (i.e. the ``direct" method). Theoretical calibrations instead make use of photoionization and stellar population synthesis models. The qualitatively different MZSRs determined from the SDSS data derived by \citet{Mannucci2010}, \citet{Yates2012} and \citet{Andrews2013} are largely due to the different abundance calibrations applied in deriving the metallicities. \citet{Yates2012} and \citet{Andrews2013} derive metallicities using purely theoretical and empirical methods, respectively. \citet{Mannucci2010} derive metallicities using a semi-empirical calibration which combines both theoretical and empirical methods.

Detailed discussions regarding the strengths and weaknesses of the various strong-line methods can be found in e.g., \citet{Kewley2008} and \citet{Moustakas2010}. Theoretical methods are not susceptible to the observational uncertainties associated with empirical methods and are capable of calibrating metallicities over the full range of observed line ratios. However, theoretical methods are model dependent and are subject the simplifying assumptions and systematic uncertainties associated with the models. In contrast, empirical methods calibrate strong-line ratios against metallicities derived from temperature sensitive auroral lines (e.g. [OIII]$\lambda4363$). The direct method provides a well understood scale for metallicity calibration. However, temperature sensitive auroral lines are extremely weak and are typically only observed in HII regions and galaxies with  $Z<Z_\odot$ and with large observational uncertainties. More to the point, the direct method is known to be susceptible to several systematic issues. \citet{Peimbert1967} points out that temperature fluctuations may lead to systematic underestimates of the metallicity. More recently, the assumption that HII regions are in thermodynamic equilibrium has also been challenged \citep{Nicholls2012, Nicholls2013, Dopita2013}. These authors argue that a breakdown of the assumption of equilibrium leads to an underestimate of the abundance, particularly in metal-rich HII regions. Furthermore, \citet{Nicholls2013} and \citet{Dopita2013} suggest that the use of old atomic data also contributes to the abundance discrepancy. Accounting for these two effects brings theoretically and empirically determined metallicities into good agreement \citep{Dopita2013}.

The suggestion that non-equilibrium electron energy distributions may be at the heart of the abundance discrepancy is promising. Astrophysical plasmas where \textit{in situ} measurements of electron energies can be made have non-equilibrium energy distributions \citep{Nicholls2012}. This suggests that perhaps metallicities determinations based on theoretical models \citep[i.e.][]{Yates2012} may be more reliable. However, given that non-equilibrium processes remain only one possible solution to the long-standing abundance discrepancy problem, the MZSR derived by \citet{Yates2012} remains uncertain. It is beyond the scope of this paper to investigate these issues in detail but in light of this uncertainty, the physical interpretation based on the observed MZSR of dust efflux as the process responsible for dust loss in the Slow Flow model remains tentative.

\citet{Lara-Lopez2013} examine the relation between stellar mass, specific-SFR and metallicity. They derive a relation similar to MZSR first shown in \citet{Yates2012}. However, \citet{Lara-Lopez2013} also examine the HI gas content. They show that for massive galaxies the gas fraction is \textit{positively} correlated with metallicity and SFR. Metallicity is a relative measure of the oxygen to hydrogen abundance in the gas-phase. Therefore, the positive correlation between metallicity, gas fraction and SFR suggests that galaxies with high metallicities and high SFRs may also have greater mass of metals. If the gas and dust are coupled, galaxies may also efflux gas and metals. The observed positive correlation between metallicity, SFR and gas fraction for massive galaxies supports the interpretation of dust efflux over dust destruction. It should be noted that \citet{Bothwell2013} have also examined the relation between stellar mass, metallicity, SFR and HI content and find similar results for less massive galaxies. However, they do not report a positive correlation between metallicity, SFR and HI gas content for massive galaxies though this may be due to systematic effects attributed to the metallicity calibration adopted in that study \citep[see][]{Yates2012}.

The dominant dust destruction mechanisms such as UV photodissociation, sputtering and shocks in star-forming galaxies are dependent on massive stars and therefore the rate of dust destruction should be temporally coincidental with the SFR. However, in order to reproduce the MDSR, a significant temporal offset is required ($\Delta t \sim 3.5$ Gyr), thus favoring an interpretation where dust is effluxed rather than destroyed. In Section 5.4, we develop a model with dust loss proportional to only the SFR (Equation \ref{eq:model_alpha}) without a temporal offset. This fast flow term can be more readily interpreted as a dust destruction term. However, we emphasize that this term is not necessary to reproduce the observed MDSR (compare Figure \ref{fig:param} and \ref{fig:param3}).

\section{Model Assumptions}

The numerical model we develop in Section 3, 4 and 5 to reproduce the observed MDSR represents one possible implementation. Here we revisit the basic galaxy properties we have assumed. These basic properties are required for any model attempting to reproduce the MDSR in accordance with the physical scenario described in Section 1.2. 

\begin{enumerate}

\item \textit{At a fixed stellar mass the current SFR is anti-correlated to the age of the galaxy.} This assumption is supported by observations which indicate that at a fixed stellar mass the mass-weighted age of stars within a galaxy is \textit{anti-}correlated to its current SFR (see Figure \ref{fig:age}). An \textit{anti-}correlation between current SFR and galaxy age is the natural consequence of a sufficiently long characteristic timescale associated with the change in the average position of galaxies within the scatter of the MSR. In an accretion driven star formation scenario, the gravitational accretion rate is modulated by dark matter halos and should scale with the halo virial time. By $z=0$ this is several Gyr which could help explain the observed anti-correlation between current SFR and galaxy age. In any case, we have implemented the observed SFR-age anti-correlation by assuming that galaxies populating the scatter of the MSR maintain a constant offset from the mean relation as they evolve. This is the easiest, albeit most restrictive, implementation to produce the \textit{anti-}correlation between galaxy age and current SFR. It is beyond the scope of this work to explore alternative approaches but we note that any implementations reproducing the \textit{anti-}correlation between galaxy age and current SFR could produce a similar model MDSR.

\item \textit{The rate of dust production is proportional to the rate at which mass is recycled by stars.} We have assumed that the dust production rate is directly proportional to the stellar mass return rate. For reproducing the MDSR the exact nature of the proportionality is not critical so long as at a fixed stellar mass the \textit{relative} quantity of dust produced is larger (smaller) for galaxies that are older (younger) and have returned more (less) stellar mass back to the ISM.

\item \textit{The rate of dust loss is dependent on the amount of dust present and the rate at which high energy photons are produced.} We have implemented this by assuming that the rate of dust loss is proportional to the product of the dust opacity and SFR (see Equation \ref{eq:model}). A dust loss rate that is given as a product of a quantity that is proportional to the high energy photon production rate and any other physical parameter that scales with the stellar mass (e.g. metallicity, dust mass or stellar mass itself) could produce a similar model MDSR.

\item \textit{The timescale of dust loss is comparable but not identical to the timescale of dust production.} The timescale of dust production is dictated by the evolutionary timescale of intermediate and high mass stars. We have taken the rate of dust production to be directly proportional to the rate of stellar mass recycling. If the dust production timescale is different from the simple assumption made, then the dust loss timescale will change commensurately. This could happen, for example, if one of the populations of stars considered the dominant producers of dust in galaxies (i.e. intermediate mass AGB or type II SN) dominates over another in terms of dust production.

\end{enumerate}
If the four physical conditions enumerated above are representative of the star-forming population of galaxies in the local universe, then a model can be developed that will reproduce the notable features of the observed MDSR shown in Figure \ref{fig:tau} and in accordance with the physical scenario described in Section 1.2. We are not aware of any observational evidence that challenges these physical conditions, though we emphasize that all the physical conditions are not well established either. The physically motivated model developed in this contribution presents an interesting falsifiable hypothesis for the origin of the MDSR.

\section{Discussion}

In Section 8.1 we discuss details of our model of stellar mass growth and in Section 8.2 we discuss outflows in galaxies. We consider some potential implications of our model in Section 8.3.

\subsection{Stellar Mass Growth and the Scatter in the Stellar Mass - SFR Relation}


The \textit{anti-}correlation between galaxy age and SFR at a fixed stellar mass is a fundamental assumption of our model. At stellar masses $\lesssim10^{10}M_\odot$ the \textit{anti-}correlation between dust opacity and SFR is a direct consequence of the \textit{anti-}correlation between age and SFR. Galaxies with low SFRs are older and therefore have had more time to produce and accumulate dust. The remarkable agreement between our model of the MDSR and the observed relation, in addition to indications from the fossil record, suggest that our model of stellar mass growth may be feasible. Thus, further investigation is warranted. The model of stellar mass growth developed in this study begs a very important question: if the scatter in the MSR is temporally correlated, what is the physical mechanism responsible for such a mode of galaxy evolution?


\subsection{Outflows in Normal Star-Forming Galaxies}

In our Slow Flow model dust mass loss scales with the opacity and SFR. Given the (few) Gyr timescales for dust mass loss inferred from our model, we favor radiation pressure acting over long periods of time as the physical basis of the model. Radiation pressure acting on dust grains has been considered both in numerical models \citep{Chiao1972, Ferrara1993, Davies1998, Murray2005, Murray2011} and cosmological simulations \citep{Aguirre2001, Hopkins2012} as a possible physical mechanism for driving outflows. However, a detailed physical model for a sustained radiation field interacting with dust grains over long periods of time incorporating some degree of dynamical coupling of dust grains to ambient gas remains to be developed. The analytical model presented in this paper provides constraints for a more rigorous physical model. 

In our model the dominant mechanism for establishing a twist in the MDSR is radiation driven outflows. A simple back-of-the-envelope estimate of the average outflow velocity can be derived by assuming a distance dust travels before it is out of the line-of-sight, $D_d$, in a time $\Delta t$. The average outflow velocity, $\bar{v}$, is then given by $\bar{v} = D_d/\Delta t$. Adopting a fiducial value of 10 kpc for $D_d$ and $\Delta t = 3.5$ Gyr determined from our model we get that $\bar{v} \sim 3$ km s$^{-1}$. If material is ejected and re-accreted back into the galaxies as some models suggest \citep{Dave2011c}, then dust may be driven out at substantially higher velocities in which case $\bar{v}$ represents a ``drift" velocity at which a net loss of dust occurs.

In the Slow Flow scenario magnetic fields may play an important role in the mass loss process. \citet{Chiao1972} consider a physical scenario of dust driven out of galaxies along magnetic field lines by radiation pressure from stars. They argue that the Larmor radius of charged dust particles in typical ISM conditions is significantly smaller than the disk thickness. Therefore, the motion of dust grains is tied to magnetic field lines. They consider the net outward force from radiation pressure balanced against the drag force from the inter-cloud medium. They calculate the drift velocity is
\begin{equation}
u = 0.82\left( \frac{0.01 \, \, \mathrm{cm}^{-3}}{n_i}\right) \left(  \frac{10^4 \, \, \mathrm{K}}{T} \right)^{1/2} \mathrm{sin} \, \theta \,\,\,\,[\mathrm{km \,\, s}^{-1}].
\end{equation}
Here $u$ is the drift velocity, $n_i$ is the number density of particles in the inter-cloud medium, $T$ is the temperature and $\theta$ is the inclination angle of the magnetic field with respect to the galactic plane. For the prevailing physical conditions in most star-forming galaxies and assuming a reasonably inclined magnetic field with respect to the galactic plane, the drift velocity of dust grains driven by radiation  pressure along magnetic field lines is on the order of a $\sim1$ km s$^{-1}$. This is on the same order of magnitude as the drift velocity inferred from our model.

In our model a large fraction of the dust produced by a galaxy can be expelled in a slow flow (see Figures \ref{fig:p2c} and \ref{fig:p3c}a). The degree to which dust is dynamically coupled to the ambient gas in the ISM of galaxies is uncertain. Both collisions of dust particles with gas or coulomb interactions of charged particles with ions could provide a mechanism for transferring momentum from the dust to the gas \citep{Ferrara1993, Davies1998, Draine2004}. If the dust and gas are strongly coupled then a significant fraction of the gas in star-forming galaxies could be expelled via the Slow Flow mechanism. In \citet{Zahid2013a} we argue that the extension of the MDSR to the quiescent population of galaxies suggests that the physical mechanism responsible for the observed MDSR may be related to the shutting down of star-formation in galaxies. The expulsion of large quantities of gas driven by interaction with charged dust particles accelerated by radiation pressure from stars presents a potential physical basis for such a scenario.

From $\sim110,000$ star-forming galaxies in the SDSS DR7, \citet{Yates2012} have observed a ``twist" in the stellar mass, metallicity and SFR relation. The twist occurs at a similar stellar mass as the observed twist in the MDSR of star-forming galaxies. Given that dust and metal content are strongly correlated, we argue that the twist in the stellar mass, metallicity and SFR relation is likely a result of the same physical mechanism responsible for the twist in the dust relation (i.e. the Slow Flow mechanism). At a stellar mass of $\sim10^{11}M_\odot$, low SFR galaxies lose $\sim50\%$ of dust they produce in the Slow Flow (see Figure \ref{fig:p2c} or \ref{fig:p3c}a). This is a factor of $\sim2$ more dust than their high SFR counterparts . Similar values for the oxygen mass loss are likely required to explain the twist observed in stellar mass, metallicity and SFR relation. For this to be the case, a relatively strong dynamical coupling between the dust and oxygen in the gas is required. 

Taken at face value, the slow flow properties inferred from our model suggest a qualitatively different type of outflow than those observed in starburst, post-starburst and luminous infrared galaxies in the nearby universe \citep{Rupke2005, Martin2006, Rich2010, Tremonti2007, Sharp2010} or star-forming galaxies at higher redshifts \citep{Shapley2003, Weiner2009, Steidel2010}. In these galaxies outflows with velocities on order $\sim1000$ km/s are observed driven by energy and/or momentum from vigorous star formation. The mass loading of the wind is typically on order unity or greater. Outflows have also been observed in the stacked spectra of normal star-forming galaxies in the local universe \citep{Chen2010}. In these normal star-forming galaxies, the outflow velocities inferred from the NaD absorption line range from $\sim120 - 160$ km s$^{-1}$ .

\citet{Hopkins2012} consider several independent feedback mechanisms in their cosmological simulations. They conclude that multiple feedback processes acting on different spatial and temporal scales are required to produce realistic outflows. In particular they find that in normal star-forming galaxies such as the Milky Way, gas is largely driven out by heating of the ISM by supernovae and shocked stellar winds. Radiation pressure does drive some cold gas out of the ISM but the fraction is small. We note that \citet{Hopkins2012} run their observations over a few orbital periods ($\sim$0.5 Gyr), future simulations run over longer timescales could provide an important test for the model presented here.

Our three parameter model suggests that there is room for additional, independent physical mechanisms for driving out (or destroying) dust in star-forming galaxies. The three parameter model incorporates an additional feedback term that is simply proportional to the SFR. This term could be physically interpreted as energy injection from SN and shocks driven by stellar winds. In this physical picture, thermal energy is deposited directly into the ISM and no coupling between dust and gas is required in order to drive outflows. It may be the case that though a substantial fraction of the \textit{dust} is lost in the slow flow, the coupling of dust and gas is weak leading to very little gas entrainment. Our three parameter model allows for $\sim30\%-40\%$ of the dust produced by galaxies at all stellar masses to be driven out by heating of the ISM (see Figure \ref{fig:p3c}b) while still producing the twist in the MDSR. The velocity and mass loading of this type of outflow could be significantly larger than the slow flow since the timescale is instantaneous and, to first-order, we expect the ambient dust-to-gas ratio ($\sim$0.01) for the outflowing material.

\subsection{Mass Transport}

Dust driven outflows offer an attractive physical mechanism to explain the dust observed outside of galaxies. Using spectropolarimetric observations, \cite{Yoshida2011} directly observe dust in the wind of M82. Additionally, the presence of dust around galaxies is established by direct observations of polycyclic aromatic hydrocarbon features \citep{Engelbracht2006, Roussel2010}, in the serendipitous alignment of a galaxy pair where extended dust disk is observed \citep{Holwerda2009} and statistical studies of extinction of background galaxies by foreground galaxy halos \citep{Zaritsky1994b, Menard2010}. Excess reddening is observed on scales ranging from a few kpc to a few Mpc. \citet{Fukugita2011} argues that the total amount of dust produced in the history of the universe is consistent with the amount observed inside and outside of galaxies suggesting that only a small fraction of dust produced is destroyed. From this analysis it is estimated that a large fraction ($\sim60\%$) of the total dust in the universe is distributed in the halos of galaxies and the intergalactic medium \citep[also see][]{Menard2012}.

In \citet{Zahid2012b} we conduct a census of oxygen in star-forming galaxies in the local universe. We find that a large fraction of oxygen produced in galaxies is not accounted for by oxygen locked up in stars and in the ISM. The ``deficit" of oxygen observed is consistent with the estimated oxygen content of the halos of star-forming galaxies \citep{Tumlinson2011}. A robust conclusion of \citet{Zahid2012b} is that the oxygen deficit scales with stellar mass such that high mass galaxies are missing a larger fraction of their oxygen as compared to lower mass galaxies. The magnitude of dust mass loss and its scaling with stellar mass presented in this study is consistent with the inferred oxygen deficit in local star-forming galaxies. The Slow Flow mechanism may provide a physical basis for understanding the oxygen census of local star-forming galaxies presented in \citet{Zahid2012b}.

The physical transport of dust and metals may help to explain several observations. In spiral galaxies, the flat oxygen abundance gradients observed out to several optical radii are inconsistent with \textit{in situ} formation of oxygen \citep{Bresolin2009a, Werk2010, Werk2011, Bresolin2012}. A plausible explanation is the transport of oxygen from the inner parts of galaxies. Such a mechanism would also be consistent with the ubiquitous presence of oxygen in the halos of star forming galaxies \citep{Tumlinson2011}.


The slow flow may also have an effect on local properties of galaxies. \citet{Bresolin2009b} show that the metallicities inferred from temperature sensitive [OIII]$\lambda$4363 line and those determined from B and A supergiants by \citet{Kudritzki2008} agree well in NGC 300. If coulomb interactions are responsible for exchange of momentum between dust and gas, then a preferential loss of more ionized species may occur in dustier HII regions. If this is the case, our model suggests that at higher metallicities, the metallicities inferred for HII regions may be systematically lower than stellar metallicities due to the efficient loss of metals from HII regions with higher dust opacities. However, this would depend on the velocities and timescales of outflows in HII regions which currently are not well constrained. Indication of this effect may already be seen in M81 \citep{Patterson2012, Kudritzki2012}, the Orion nebula \citep{Simon-Diaz2011}, in the central region of M33 \citep{Urbaneja2005, U2009, Bresolin2010} and in M31 \citep{Zurita2012}. 

\section{Summary}


We develop a physically motivated model to reproduce and explain the observed relation between stellar mass, dust opacity and SFR (MDSR). In our model, galaxies continuously evolve along the observed MSR. The scatter in the MSR is populated by galaxies in a temporally correlated manner. As they build up their stellar mass, they produce dust and become more opaque. Radiation from massive stars efficiently couples to dust in systems with high opacity and over long periods of time slowly drives out dust and presumably gas that may be dynamically coupled to the dust. Models for stellar mass growth whereby star-forming galaxies continuously evolve along the MSR and models of radiation pressure acting on dust grains have both been investigated in the literature. This paper is the first to put these together in a self-consistent model. We refer to this self-consistent model incorporating both stellar mass growth and radiation pressure driven outflows, given by Equation \ref{eq:scale_model}, as the \textit{Slow Flow} model. We summarize the salient features of our model:

\begin{itemize}
\item Normal star-forming galaxies evolve along the observed stellar mass-SFR relation. The $\sim0.25$ dex $1\sigma$ scatter in the relation is temporally correlated for individual galaxies. Thus, for galaxies at the same stellar mass, the age of a galaxy is \textit{anti-}correlated with SFR such that older galaxies have lower \textit{current} SFRs.

\item Stellar mass recycling is time dependent and the dust production rate is proportional to the stellar mass recycling rate. At the fixed stellar mass, older galaxies have recycled a larger fraction of their gas back to the ISM and thus produced a greater amount of dust. Also, older galaxies have lower current SFRs at a fixed stellar mass. Therefore, the \textit{anti-}correlation between dust opacity and SFR at stellar masses $\lesssim10^{10}M_\odot$ is naturally explained as an age effect.

\item Radiation pressure acting on dust grains over long periods is capable of driving dust out of the ISM of star-forming galaxies. As galaxies grow in stellar mass they become dustier. At a stellar mass $\sim10^{10}M_\odot$ star-forming galaxies become optically thick and radiation efficiently drives dust out of galaxies on a few Gyr timescales.

\item The dust production rate is proportional to the stellar recycling rate and thus is \textit{positively} correlated with the SFR. The timescale for dust loss is long and therefore at stellar masses $\gtrsim10^{10}M_\odot$, galaxies with high SFRs accumulate dust more rapidly than they expel it. The accumulation of dust in rapidly star-forming galaxies naturally explains the \textit{positive} correlation between dust opacity and SFR at stellar masses $\gtrsim10^{10}M_\odot$.

\end{itemize}

The results presented in this paper offer interesting new evidence that outflows of gas and dust play an important role in the evolution of normal star-forming galaxies in the local universe. We demonstrate that considerations for time dependent stellar mass loss are important for understanding the properties of star-forming galaxies. Despite being quite simple, our model gives us a potentially powerful tool for interpreting the observed relation between stellar mass, dust opacity and SFR. A detailed model of the physical processes outlined in this paper needs to be worked out and observations of outflows in normal star-forming galaxies are required to establish and constrain the model.

\vspace{1 cm}
We thank the anonymous reviewer for carefully reading the manuscript and providing constructive comments. This paper benefitted from useful discussions with Brett Andrews, Charlie Conroy, Lars Hernquist, Phil Hopkins, Josh Barnes and Jessica Werk. We are grateful to Rita Tojeiro for help with the VESPA catalogue. HJZ and LJK gratefully acknowledge support by NSF EARLY CAREER AWARD AST07-48559. HJZ enjoyed the hospitality of Margaret Geller and the Smithsonian Astrophysical Observatory during the completion of this work. RPK acknowledges support by the Alexander-von-Humboldt Foundation and the hospitality of the Max-Planck-Institute for Astrophysics in Garching where part of this work was carried out.

 

\bibliographystyle{apj}
\bibliography{metallicity}

\begin{thebibliography}{116}
\expandafter\ifx\csname natexlab\endcsname\relax\def\natexlab#1{#1}\fi

\bibitem[{{Abazajian} {et~al.}(2009){Abazajian}, {Adelman-McCarthy},
  {Ag{\"u}eros}, {Allam}, {Allende Prieto}, {An}, {Anderson}, {Anderson},
  {Annis}, {Bahcall}, \& et~al.}]{Abazajian2009}
{Abazajian}, K.~N., {et~al.} 2009, \apjs, 182, 543

\bibitem[{{Aguirre} {et~al.}(2001){Aguirre}, {Hernquist}, {Katz}, {Gardner}, \&
  {Weinberg}}]{Aguirre2001}
{Aguirre}, A., {Hernquist}, L., {Katz}, N., {Gardner}, J., \& {Weinberg}, D.
  2001, \apjl, 556, L11

\bibitem[{{Andrews} \& {Martini}(2013)}]{Andrews2013}
{Andrews}, B.~H., \& {Martini}, P. 2013, \apj, 765, 140

\bibitem[{{Asari} {et~al.}(2007){Asari}, {Cid Fernandes}, {Stasi{\'n}ska},
  {Torres-Papaqui}, {Mateus}, {Sodr{\'e}}, {Schoenell}, \& {Gomes}}]{Asari2007}
{Asari}, N.~V., {Cid Fernandes}, R., {Stasi{\'n}ska}, G., {Torres-Papaqui},
  J.~P., {Mateus}, A., {Sodr{\'e}}, L., {Schoenell}, W., \& {Gomes}, J.~M.
  2007, \mnras, 381, 263

\bibitem[{{Berg} {et~al.}(2012){Berg}, {Skillman}, {Marble}, {van Zee},
  {Engelbracht}, {Lee}, {Kennicutt}, {Calzetti}, {Dale}, \&
  {Johnson}}]{Berg2012}
{Berg}, D.~A., {et~al.} 2012, \apj, 754, 98

\bibitem[{{Boissier} {et~al.}(2004){Boissier}, {Boselli}, {Buat}, {Donas}, \&
  {Milliard}}]{Boissier2004}
{Boissier}, S., {Boselli}, A., {Buat}, V., {Donas}, J., \& {Milliard}, B. 2004,
  \aap, 424, 465

\bibitem[{{Bothwell} {et~al.}(2013){Bothwell}, {Maiolino}, {Kennicutt},
  {Cresci}, {Mannucci}, {Marconi}, \& {Cicone}}]{Bothwell2013}
{Bothwell}, M.~S., {Maiolino}, R., {Kennicutt}, R., {Cresci}, G., {Mannucci},
  F., {Marconi}, A., \& {Cicone}, C. 2013, \mnras, 433, 1425

\bibitem[{{Bresolin} {et~al.}(2004){Bresolin}, {Garnett}, \&
  {Kennicutt}}]{Bresolin2004}
{Bresolin}, F., {Garnett}, D.~R., \& {Kennicutt}, Jr., R.~C. 2004, \apj, 615,
  228

\bibitem[{{Bresolin} {et~al.}(2009{\natexlab{a}}){Bresolin}, {Gieren},
  {Kudritzki}, {Pietrzy{\'n}ski}, {Urbaneja}, \& {Carraro}}]{Bresolin2009b}
{Bresolin}, F., {Gieren}, W., {Kudritzki}, R.-P., {Pietrzy{\'n}ski}, G.,
  {Urbaneja}, M.~A., \& {Carraro}, G. 2009{\natexlab{a}}, \apj, 700, 309

\bibitem[{{Bresolin} {et~al.}(2012){Bresolin}, {Kennicutt}, \&
  {Ryan-Weber}}]{Bresolin2012}
{Bresolin}, F., {Kennicutt}, R.~C., \& {Ryan-Weber}, E. 2012, \apj, 750, 122

\bibitem[{{Bresolin} {et~al.}(2009{\natexlab{b}}){Bresolin}, {Ryan-Weber},
  {Kennicutt}, \& {Goddard}}]{Bresolin2009a}
{Bresolin}, F., {Ryan-Weber}, E., {Kennicutt}, R.~C., \& {Goddard}, Q.
  2009{\natexlab{b}}, \apj, 695, 580

\bibitem[{{Bresolin} {et~al.}(2010){Bresolin}, {Stasi{\'n}ska},
  {V{\'{\i}}lchez}, {Simon}, \& {Rosolowsky}}]{Bresolin2010}
{Bresolin}, F., {Stasi{\'n}ska}, G., {V{\'{\i}}lchez}, J.~M., {Simon}, J.~D.,
  \& {Rosolowsky}, E. 2010, \mnras, 404, 1679

\bibitem[{{Brinchmann} {et~al.}(2004){Brinchmann}, {Charlot}, {White},
  {Tremonti}, {Kauffmann}, {Heckman}, \& {Brinkmann}}]{Brinchmann2004}
{Brinchmann}, J., {Charlot}, S., {White}, S.~D.~M., {Tremonti}, C.,
  {Kauffmann}, G., {Heckman}, T., \& {Brinkmann}, J. 2004, \mnras, 351, 1151

\bibitem[{{Cardelli} {et~al.}(1989){Cardelli}, {Clayton}, \&
  {Mathis}}]{Cardelli1989}
{Cardelli}, J.~A., {Clayton}, G.~C., \& {Mathis}, J.~S. 1989, \apj, 345, 245

\bibitem[{{Chabrier}(2003)}]{Chabrier2003}
{Chabrier}, G. 2003, \pasp, 115, 763

\bibitem[{{Chattopadhyay} {et~al.}(2012){Chattopadhyay}, {Sharma}, {Nath}, \&
  {Ryu}}]{Chattopadhyay2012}
{Chattopadhyay}, I., {Sharma}, M., {Nath}, B.~B., \& {Ryu}, D. 2012, \mnras,
  423, 2153

\bibitem[{{Chen} {et~al.}(2010){Chen}, {Tremonti}, {Heckman}, {Kauffmann},
  {Weiner}, {Brinchmann}, \& {Wang}}]{Chen2010}
{Chen}, Y.-M., {Tremonti}, C.~A., {Heckman}, T.~M., {Kauffmann}, G., {Weiner},
  B.~J., {Brinchmann}, J., \& {Wang}, J. 2010, \aj, 140, 445

\bibitem[{{Chiao} \& {Wickramasinghe}(1972)}]{Chiao1972}
{Chiao}, R.~Y., \& {Wickramasinghe}, N.~C. 1972, \mnras, 159, 361

\bibitem[{{Conroy} \& {Wechsler}(2009)}]{Conroy2009b}
{Conroy}, C., \& {Wechsler}, R.~H. 2009, \apj, 696, 620

\bibitem[{{Cowie} \& {Barger}(2008)}]{Cowie2008}
{Cowie}, L.~L., \& {Barger}, A.~J. 2008, \apj, 686, 72

\bibitem[{{Daddi} {et~al.}(2007){Daddi}, {Dickinson}, {Morrison}, {Chary},
  {Cimatti}, {Elbaz}, {Frayer}, {Renzini}, {Pope}, {Alexander}, {Bauer},
  {Giavalisco}, {Huynh}, {Kurk}, \& {Mignoli}}]{Daddi2007}
{Daddi}, E., {et~al.} 2007, \apj, 670, 156

\bibitem[{{Dav{\'e}} {et~al.}(2011){Dav{\'e}}, {Finlator}, \&
  {Oppenheimer}}]{Dave2011c}
{Dav{\'e}}, R., {Finlator}, K., \& {Oppenheimer}, B.~D. 2011, \mnras, 2110

\bibitem[{{Davies} {et~al.}(1998){Davies}, {Alton}, {Bianchi}, \&
  {Trewhella}}]{Davies1998}
{Davies}, J.~I., {Alton}, P., {Bianchi}, S., \& {Trewhella}, M. 1998, \mnras,
  300, 1006

\bibitem[{{Dom{\'{\i}}nguez} {et~al.}(2013){Dom{\'{\i}}nguez}, {Siana},
  {Henry}, {Scarlata}, {Bedregal}, {Malkan}, {Atek}, {Ross}, {Colbert},
  {Teplitz}, {Rafelski}, {McCarthy}, {Bunker}, {Hathi}, {Dressler}, {Martin},
  \& {Masters}}]{Dominguez2013}
{Dom{\'{\i}}nguez}, A., {et~al.} 2013, \apj, 763, 145

\bibitem[{{Dopita} {et~al.}(2013){Dopita}, {Sutherland}, {Nicholls}, {Kewley},
  \& {Vogt}}]{Dopita2013}
{Dopita}, M.~A., {Sutherland}, R.~S., {Nicholls}, D.~C., {Kewley}, L.~J., \&
  {Vogt}, F.~P.~A. 2013, ArXiv e-prints

\bibitem[{{Draine}(2003)}]{Draine2003}
{Draine}, B.~T. 2003, \araa, 41, 241

\bibitem[{{Draine}(2004)}]{Draine2004}
---. 2004, {Astrophysics of Dust in Cold Clouds} (Springer), 213

\bibitem[{{Dutton} {et~al.}(2010){Dutton}, {van den Bosch}, \&
  {Dekel}}]{Dutton2010}
{Dutton}, A.~A., {van den Bosch}, F.~C., \& {Dekel}, A. 2010, \mnras, 405, 1690

\bibitem[{{Dwek}(1998)}]{Dwek1998}
{Dwek}, E. 1998, \apj, 501, 643

\bibitem[{{Elbaz} {et~al.}(2007){Elbaz}, {Daddi}, {Le Borgne}, {Dickinson},
  {Alexander}, {Chary}, {Starck}, {Brandt}, {Kitzbichler}, {MacDonald},
  {Nonino}, {Popesso}, {Stern}, \& {Vanzella}}]{Elbaz2007}
{Elbaz}, D., {et~al.} 2007, \aap, 468, 33

\bibitem[{{Ellison} {et~al.}(2008){Ellison}, {Patton}, {Simard}, \&
  {McConnachie}}]{Ellison2008}
{Ellison}, S.~L., {Patton}, D.~R., {Simard}, L., \& {McConnachie}, A.~W. 2008,
  \apjl, 672, L107

\bibitem[{{Engelbracht} {et~al.}(2006){Engelbracht}, {Kundurthy}, {Gordon},
  {Rieke}, {Kennicutt}, {Smith}, {Regan}, {Makovoz}, {Sosey}, {Draine},
  {Helou}, {Armus}, {Calzetti}, {Meyer}, {Bendo}, {Walter}, {Hollenbach},
  {Cannon}, {Murphy}, {Dale}, {Buckalew}, \& {Sheth}}]{Engelbracht2006}
{Engelbracht}, C.~W., {et~al.} 2006, \apjl, 642, L127

\bibitem[{{Erb} {et~al.}(2006){Erb}, {Shapley}, {Pettini}, {Steidel}, {Reddy},
  \& {Adelberger}}]{Erb2006b}
{Erb}, D.~K., {Shapley}, A.~E., {Pettini}, M., {Steidel}, C.~C., {Reddy},
  N.~A., \& {Adelberger}, K.~L. 2006, \apj, 644, 813

\bibitem[{{Ferrara}(1993)}]{Ferrara1993}
{Ferrara}, A. 1993, \apj, 407, 157

\bibitem[{{Fukugita}(2011)}]{Fukugita2011}
{Fukugita}, M. 2011, ArXiv e-prints

\bibitem[{{Garn} \& {Best}(2010)}]{Garn2010b}
{Garn}, T., \& {Best}, P.~N. 2010, \mnras, 409, 421

\bibitem[{{Heckman} {et~al.}(1998){Heckman}, {Robert}, {Leitherer}, {Garnett},
  \& {van der Rydt}}]{Heckman1998}
{Heckman}, T.~M., {Robert}, C., {Leitherer}, C., {Garnett}, D.~R., \& {van der
  Rydt}, F. 1998, \apj, 503, 646

\bibitem[{{Holwerda} {et~al.}(2009){Holwerda}, {Keel}, {Williams}, {Dalcanton},
  \& {de Jong}}]{Holwerda2009}
{Holwerda}, B.~W., {Keel}, W.~C., {Williams}, B., {Dalcanton}, J.~J., \& {de
  Jong}, R.~S. 2009, \aj, 137, 3000

\bibitem[{{Hopkins} {et~al.}(2012){Hopkins}, {Quataert}, \&
  {Murray}}]{Hopkins2012}
{Hopkins}, P.~F., {Quataert}, E., \& {Murray}, N. 2012, \mnras, 421, 3522

\bibitem[{{Jungwiert} {et~al.}(2001){Jungwiert}, {Combes}, \& {Palou{\v
  s}}}]{Jungwiert2001}
{Jungwiert}, B., {Combes}, F., \& {Palou{\v s}}, J. 2001, \aap, 376, 85

\bibitem[{{Kauffmann} {et~al.}(2003{\natexlab{a}}){Kauffmann}, {Heckman},
  {Tremonti}, {Brinchmann}, {Charlot}, {White}, {Ridgway}, {Brinkmann},
  {Fukugita}, {Hall}, {Ivezi{\'c}}, {Richards}, \& {Schneider}}]{Kauffmann2003}
{Kauffmann}, G., {et~al.} 2003{\natexlab{a}}, \mnras, 346, 1055

\bibitem[{{Kauffmann} {et~al.}(2003{\natexlab{b}}){Kauffmann}, {Heckman},
  {White}, {Charlot}, {Tremonti}, {Brinchmann}, {Bruzual}, {Peng}, {Seibert},
  {Bernardi}, {Blanton}, {Brinkmann}, {Castander}, {Cs{\'a}bai}, {Fukugita},
  {Ivezic}, {Munn}, {Nichol}, {Padmanabhan}, {Thakar}, {Weinberg}, \&
  {York}}]{Kauffmann2003a}
---. 2003{\natexlab{b}}, \mnras, 341, 33

\bibitem[{{Kennicutt} {et~al.}(2003){Kennicutt}, {Bresolin}, \&
  {Garnett}}]{Kennicutt2003}
{Kennicutt}, Jr., R.~C., {Bresolin}, F., \& {Garnett}, D.~R. 2003, \apj, 591,
  801

\bibitem[{{Kewley} \& {Ellison}(2008)}]{Kewley2008}
{Kewley}, L.~J., \& {Ellison}, S.~L. 2008, \apj, 681, 1183

\bibitem[{{Kewley} {et~al.}(2006){Kewley}, {Groves}, {Kauffmann}, \&
  {Heckman}}]{Kewley2006}
{Kewley}, L.~J., {Groves}, B., {Kauffmann}, G., \& {Heckman}, T. 2006, \mnras,
  372, 961

\bibitem[{{Kobulnicky} \& {Kewley}(2004)}]{Kobulnicky2004}
{Kobulnicky}, H.~A., \& {Kewley}, L.~J. 2004, \apj, 617, 240

\bibitem[{{Koyama} {et~al.}(2013){Koyama}, {Smail}, {Kurk}, {Geach}, {Sobral},
  {Kodama}, {Nakata}, {Swinbank}, {Best}, {Hayashi}, \& {Tadaki}}]{Koyama2013}
{Koyama}, Y., {et~al.} 2013, \mnras

\bibitem[{{Kudritzki} {et~al.}(2008){Kudritzki}, {Urbaneja}, {Bresolin},
  {Przybilla}, {Gieren}, \& {Pietrzy{\'n}ski}}]{Kudritzki2008}
{Kudritzki}, R.-P., {Urbaneja}, M.~A., {Bresolin}, F., {Przybilla}, N.,
  {Gieren}, W., \& {Pietrzy{\'n}ski}, G. 2008, \apj, 681, 269

\bibitem[{{Kudritzki} {et~al.}(2012){Kudritzki}, {Urbaneja}, {Gazak},
  {Bresolin}, {Przybilla}, {Gieren}, \& {Pietrzy{\'n}ski}}]{Kudritzki2012}
{Kudritzki}, R.-P., {Urbaneja}, M.~A., {Gazak}, Z., {Bresolin}, F.,
  {Przybilla}, N., {Gieren}, W., \& {Pietrzy{\'n}ski}, G. 2012, \apj, 747, 15

\bibitem[{{Lara-L{\'o}pez} {et~al.}(2010){Lara-L{\'o}pez}, {Cepa},
  {Bongiovanni}, {P{\'e}rez Garc{\'{\i}}a}, {Ederoclite}, {Casta{\~n}eda},
  {Fern{\'a}ndez Lorenzo}, {Povi{\'c}}, \&
  {S{\'a}nchez-Portal}}]{Lara-Lopez2010}
{Lara-L{\'o}pez}, M.~A., {et~al.} 2010, \aap, 521, L53+

\bibitem[{{Lara-L{\'o}pez} {et~al.}(2013){Lara-L{\'o}pez}, {Hopkins},
  {L{\'o}pez-S{\'a}nchez}, {Brough}, {Bland-Hawthorn}, {Driver}, {Foster},
  {Liske}, {Loveday}, {Robotham}, {Sharp}, {Steele}, \&
  {Taylor}}]{Lara-Lopez2013}
---. 2013, ArXiv e-prints

\bibitem[{{Laskar} {et~al.}(2011){Laskar}, {Berger}, \& {Chary}}]{Laskar2011}
{Laskar}, T., {Berger}, E., \& {Chary}, R.-R. 2011, \apj, 739, 1

\bibitem[{{Lee} {et~al.}(2006){Lee}, {Skillman}, {Cannon}, {Jackson}, {Gehrz},
  {Polomski}, \& {Woodward}}]{Lee2006}
{Lee}, H., {Skillman}, E.~D., {Cannon}, J.~M., {Jackson}, D.~C., {Gehrz},
  R.~D., {Polomski}, E.~F., \& {Woodward}, C.~E. 2006, \apj, 647, 970

\bibitem[{{Leitner}(2012)}]{Leitner2012}
{Leitner}, S.~N. 2012, \apj, 745, 149

\bibitem[{{Leitner} \& {Kravtsov}(2011)}]{Leitner2011}
{Leitner}, S.~N., \& {Kravtsov}, A.~V. 2011, \apj, 734, 48

\bibitem[{{Lequeux} {et~al.}(1979){Lequeux}, {Peimbert}, {Rayo}, {Serrano}, \&
  {Torres-Peimbert}}]{Lequeux1979}
{Lequeux}, J., {Peimbert}, M., {Rayo}, J.~F., {Serrano}, A., \&
  {Torres-Peimbert}, S. 1979, \aap, 80, 155

\bibitem[{{Mannucci} {et~al.}(2010){Mannucci}, {Cresci}, {Maiolino}, {Marconi},
  \& {Gnerucci}}]{Mannucci2010}
{Mannucci}, F., {Cresci}, G., {Maiolino}, R., {Marconi}, A., \& {Gnerucci}, A.
  2010, \mnras, 408, 2115

\bibitem[{{Mannucci} {et~al.}(2009){Mannucci}, {Cresci}, {Maiolino}, {Marconi},
  {Pastorini}, {Pozzetti}, {Gnerucci}, {Risaliti}, {Schneider}, {Lehnert}, \&
  {Salvati}}]{Mannucci2009}
{Mannucci}, F., {et~al.} 2009, \mnras, 398, 1915

\bibitem[{{Maraston}(2005)}]{Maraston2005}
{Maraston}, C. 2005, \mnras, 362, 799

\bibitem[{{Martin}(2006)}]{Martin2006}
{Martin}, C.~L. 2006, \apj, 647, 222

\bibitem[{{M{\'e}nard} \& {Fukugita}(2012)}]{Menard2012}
{M{\'e}nard}, B., \& {Fukugita}, M. 2012, \apj, 754, 116

\bibitem[{{M{\'e}nard} {et~al.}(2010){M{\'e}nard}, {Scranton}, {Fukugita}, \&
  {Richards}}]{Menard2010}
{M{\'e}nard}, B., {Scranton}, R., {Fukugita}, M., \& {Richards}, G. 2010,
  \mnras, 405, 1025

\bibitem[{{Moustakas} {et~al.}(2010){Moustakas}, {Kennicutt}, {Tremonti},
  {Dale}, {Smith}, \& {Calzetti}}]{Moustakas2010}
{Moustakas}, J., {Kennicutt}, Jr., R.~C., {Tremonti}, C.~A., {Dale}, D.~A.,
  {Smith}, J.-D.~T., \& {Calzetti}, D. 2010, \apjs, 190, 233

\bibitem[{{Moustakas} {et~al.}(2011){Moustakas}, {Zaritsky}, {Brown}, {Cool},
  {Dey}, {Eisenstein}, {Gonzalez}, {Jannuzi}, {Jones}, {Kochanek}, {Murray}, \&
  {Wild}}]{Moustakas2011}
{Moustakas}, J., {et~al.} 2011, ArXiv e-prints

\bibitem[{{Murray} {et~al.}(2011){Murray}, {M{\'e}nard}, \&
  {Thompson}}]{Murray2011}
{Murray}, N., {M{\'e}nard}, B., \& {Thompson}, T.~A. 2011, \apj, 735, 66

\bibitem[{{Murray} {et~al.}(2005){Murray}, {Quataert}, \&
  {Thompson}}]{Murray2005}
{Murray}, N., {Quataert}, E., \& {Thompson}, T.~A. 2005, \apj, 618, 569

\bibitem[{{Nagao} {et~al.}(2006){Nagao}, {Maiolino}, \& {Marconi}}]{Nagao2006}
{Nagao}, T., {Maiolino}, R., \& {Marconi}, A. 2006, \aap, 459, 85

\bibitem[{{Nicholls} {et~al.}(2012){Nicholls}, {Dopita}, \&
  {Sutherland}}]{Nicholls2012}
{Nicholls}, D.~C., {Dopita}, M.~A., \& {Sutherland}, R.~S. 2012, \apj, 752, 148

\bibitem[{{Nicholls} {et~al.}(2013){Nicholls}, {Dopita}, {Sutherland},
  {Kewley}, \& {Palay}}]{Nicholls2013}
{Nicholls}, D.~C., {Dopita}, M.~A., {Sutherland}, R.~S., {Kewley}, L.~J., \&
  {Palay}, E. 2013, \apjs, 207, 21

\bibitem[{{Noeske} {et~al.}(2007{\natexlab{a}}){Noeske}, {Faber}, {Weiner},
  {Koo}, {Primack}, {Dekel}, {Papovich}, {Conselice}, {Le Floc'h}, {Rieke},
  {Coil}, {Lotz}, {Somerville}, \& {Bundy}}]{Noeske2007b}
{Noeske}, K.~G., {et~al.} 2007{\natexlab{a}}, \apjl, 660, L47

\bibitem[{{Noeske} {et~al.}(2007{\natexlab{b}}){Noeske}, {Weiner}, {Faber},
  {Papovich}, {Koo}, {Somerville}, {Bundy}, {Conselice}, {Newman},
  {Schiminovich}, {Le Floc'h}, {Coil}, {Rieke}, {Lotz}, {Primack}, {Barmby},
  {Cooper}, {Davis}, {Ellis}, {Fazio}, {Guhathakurta}, {Huang}, {Kassin},
  {Martin}, {Phillips}, {Rich}, {Small}, {Willmer}, \& {Wilson}}]{Noeske2007a}
---. 2007{\natexlab{b}}, \apjl, 660, L43

\bibitem[{{Osterbrock}(1989)}]{Osterbrock1989}
{Osterbrock}, D.~E. 1989, {Astrophysics of gaseous nebulae and active galactic
  nuclei}, ed. D.~E. {Osterbrock}

\bibitem[{{Pannella} {et~al.}(2009){Pannella}, {Carilli}, {Daddi}, {McCracken},
  {Owen}, {Renzini}, {Strazzullo}, {Civano}, {Koekemoer}, {Schinnerer},
  {Scoville}, {Smol{\v c}i{\'c}}, {Taniguchi}, {Aussel}, {Kneib}, {Ilbert},
  {Mellier}, {Salvato}, {Thompson}, \& {Willott}}]{Pannella2009}
{Pannella}, M., {et~al.} 2009, \apjl, 698, L116

\bibitem[{{Patterson} {et~al.}(2012){Patterson}, {Walterbos}, {Kennicutt},
  {Chiappini}, \& {Thilker}}]{Patterson2012}
{Patterson}, M.~T., {Walterbos}, R.~A.~M., {Kennicutt}, R.~C., {Chiappini}, C.,
  \& {Thilker}, D.~A. 2012, \mnras, 422, 401

\bibitem[{{Peimbert}(1967)}]{Peimbert1967}
{Peimbert}, M. 1967, \apj, 150, 825

\bibitem[{{Peng} {et~al.}(2010){Peng}, {Lilly}, {Kova{\v c}}, {Bolzonella},
  {Pozzetti}, {Renzini}, {Zamorani}, {Ilbert}, {Knobel}, {Iovino}, {Maier},
  {Cucciati}, {Tasca}, {Carollo}, {Silverman}, {Kampczyk}, {de Ravel},
  {Sanders}, {Scoville}, {Contini}, {Mainieri}, {Scodeggio}, {Kneib}, {Le
  F{\`e}vre}, {Bardelli}, {Bongiorno}, {Caputi}, {Coppa}, {de la Torre},
  {Franzetti}, {Garilli}, {Lamareille}, {Le Borgne}, {Le Brun}, {Mignoli},
  {Perez Montero}, {Pello}, {Ricciardelli}, {Tanaka}, {Tresse}, {Vergani},
  {Welikala}, {Zucca}, {Oesch}, {Abbas}, {Barnes}, {Bordoloi}, {Bottini},
  {Cappi}, {Cassata}, {Cimatti}, {Fumana}, {Hasinger}, {Koekemoer},
  {Leauthaud}, {Maccagni}, {Marinoni}, {McCracken}, {Memeo}, {Meneux}, {Nair},
  {Porciani}, {Presotto}, \& {Scaramella}}]{Peng2010}
{Peng}, Y.-j., {et~al.} 2010, \apj, 721, 193

\bibitem[{{Reddy} {et~al.}(2010){Reddy}, {Erb}, {Pettini}, {Steidel}, \&
  {Shapley}}]{Reddy2010}
{Reddy}, N.~A., {Erb}, D.~K., {Pettini}, M., {Steidel}, C.~C., \& {Shapley},
  A.~E. 2010, \apj, 712, 1070

\bibitem[{{Reddy} {et~al.}(2012){Reddy}, {Pettini}, {Steidel}, {Shapley},
  {Erb}, \& {Law}}]{Reddy2012}
{Reddy}, N.~A., {Pettini}, M., {Steidel}, C.~C., {Shapley}, A.~E., {Erb},
  D.~K., \& {Law}, D.~R. 2012, \apj, 754, 25

\bibitem[{{Rich} {et~al.}(2010){Rich}, {Dopita}, {Kewley}, \&
  {Rupke}}]{Rich2010}
{Rich}, J.~A., {Dopita}, M.~A., {Kewley}, L.~J., \& {Rupke}, D.~S.~N. 2010,
  \apj, 721, 505

\bibitem[{{Roussel} {et~al.}(2010){Roussel}, {Wilson}, {Vigroux}, {Isaak},
  {Sauvage}, {Madden}, {Auld}, {Baes}, {Barlow}, {Bendo}, {Bock}, {Boselli},
  {Bradford}, {Buat}, {Castro-Rodriguez}, {Chanial}, {Charlot}, {Ciesla},
  {Clements}, {Cooray}, {Cormier}, {Cortese}, {Davies}, {Dwek}, {Eales},
  {Elbaz}, {Galametz}, {Galliano}, {Gear}, {Glenn}, {Gomez}, {Griffin}, {Hony},
  {Levenson}, {Lu}, {O'Halloran}, {Okumura}, {Oliver}, {Page}, {Panuzzo},
  {Papageorgiou}, {Parkin}, {Perez-Fournon}, {Pohlen}, {Rangwala}, {Rigby},
  {Rykala}, {Sacchi}, {Schulz}, {Schirm}, {Smith}, {Spinoglio}, {Stevens},
  {Srinivasan}, {Symeonidis}, {Trichas}, {Vaccari}, {Wozniak}, {Wright}, \&
  {Zeilinger}}]{Roussel2010}
{Roussel}, H., {et~al.} 2010, \aap, 518, L66

\bibitem[{{Rupke} {et~al.}(2005){Rupke}, {Veilleux}, \& {Sanders}}]{Rupke2005}
{Rupke}, D.~S., {Veilleux}, S., \& {Sanders}, D.~B. 2005, \apjs, 160, 115

\bibitem[{{Salim} {et~al.}(2007){Salim}, {Rich}, {Charlot}, {Brinchmann},
  {Johnson}, {Schiminovich}, {Seibert}, {Mallery}, {Heckman}, {Forster},
  {Friedman}, {Martin}, {Morrissey}, {Neff}, {Small}, {Wyder}, {Bianchi},
  {Donas}, {Lee}, {Madore}, {Milliard}, {Szalay}, {Welsh}, \& {Yi}}]{Salim2007}
{Salim}, S., {et~al.} 2007, \apjs, 173, 267

\bibitem[{{Savaglio} {et~al.}(2005){Savaglio}, {Glazebrook}, {Le Borgne},
  {Juneau}, {Abraham}, {Chen}, {Crampton}, {McCarthy}, {Carlberg}, {Marzke},
  {Roth}, {J{\o}rgensen}, \& {Murowinski}}]{Savaglio2005}
{Savaglio}, S., {et~al.} 2005, \apj, 635, 260

\bibitem[{{Shapley} {et~al.}(2003){Shapley}, {Steidel}, {Pettini}, \&
  {Adelberger}}]{Shapley2003}
{Shapley}, A.~E., {Steidel}, C.~C., {Pettini}, M., \& {Adelberger}, K.~L. 2003,
  \apj, 588, 65

\bibitem[{{Sharma} \& {Nath}(2012)}]{Sharma2012}
{Sharma}, M., \& {Nath}, B.~B. 2012, \apj, 750, 55

\bibitem[{{Sharma} {et~al.}(2011){Sharma}, {Nath}, \&
  {Shchekinov}}]{Sharma2011}
{Sharma}, M., {Nath}, B.~B., \& {Shchekinov}, Y. 2011, \apjl, 736, L27

\bibitem[{{Sharp} \& {Bland-Hawthorn}(2010)}]{Sharp2010}
{Sharp}, R.~G., \& {Bland-Hawthorn}, J. 2010, \apj, 711, 818

\bibitem[{{Sim{\'o}n-D{\'{\i}}az} \& {Stasi{\'n}ska}(2011)}]{Simon-Diaz2011}
{Sim{\'o}n-D{\'{\i}}az}, S., \& {Stasi{\'n}ska}, G. 2011, \aap, 526, A48

\bibitem[{{Steidel} {et~al.}(2010){Steidel}, {Erb}, {Shapley}, {Pettini},
  {Reddy}, {Bogosavljevi{\'c}}, {Rudie}, \& {Rakic}}]{Steidel2010}
{Steidel}, C.~C., {Erb}, D.~K., {Shapley}, A.~E., {Pettini}, M., {Reddy}, N.,
  {Bogosavljevi{\'c}}, M., {Rudie}, G.~C., \& {Rakic}, O. 2010, \apj, 717, 289

\bibitem[{{Stoughton} {et~al.}(2002){Stoughton}, {Lupton}, {Bernardi},
  {Blanton}, {Burles}, {Castander}, {Connolly}, {Eisenstein}, {Frieman},
  {Hennessy}, {Hindsley}, {Ivezi{\'c}}, {Kent}, {Kunszt}, {Lee}, {Meiksin},
  {Munn}, {Newberg}, {Nichol}, {Nicinski}, {Pier}, {Richards}, {Richmond},
  {Schlegel}, {Smith}, {Strauss}, {SubbaRao}, {Szalay}, {Thakar}, {Tucker},
  {Vanden Berk}, {Yanny}, {Adelman}, {Anderson}, {Anderson}, {Annis},
  {Bahcall}, {Bakken}, {Bartelmann}, {Bastian}, {Bauer}, {Berman},
  {B{\"o}hringer}, {Boroski}, {Bracker}, {Briegel}, {Briggs}, {Brinkmann},
  {Brunner}, {Carey}, {Carr}, {Chen}, {Christian}, {Colestock}, {Crocker},
  {Csabai}, {Czarapata}, {Dalcanton}, {Davidsen}, {Davis}, {Dehnen},
  {Dodelson}, {Doi}, {Dombeck}, {Donahue}, {Ellman}, {Elms}, {Evans}, {Eyer},
  {Fan}, {Federwitz}, {Friedman}, {Fukugita}, {Gal}, {Gillespie}, {Glazebrook},
  {Gray}, {Grebel}, {Greenawalt}, {Greene}, {Gunn}, {de Haas}, {Haiman},
  {Haldeman}, {Hall}, {Hamabe}, {Hansen}, {Harris}, {Harris}, {Harvanek},
  {Hawley}, {Hayes}, {Heckman}, {Helmi}, {Henden}, {Hogan}, {Hogg}, {Holmgren},
  {Holtzman}, {Huang}, {Hull}, {Ichikawa}, {Ichikawa}, {Johnston}, {Kauffmann},
  {Kim}, {Kimball}, {Kinney}, {Klaene}, {Kleinman}, {Klypin}, {Knapp},
  {Korienek}, {Krolik}, {Kron}, {Krzesi{\'n}ski}, {Lamb}, {Leger},
  {Limmongkol}, {Lindenmeyer}, {Long}, {Loomis}, {Loveday}, {MacKinnon},
  {Mannery}, {Mantsch}, {Margon}, {McGehee}, {McKay}, {McLean}, {Menou},
  {Merelli}, {Mo}, {Monet}, {Nakamura}, {Narayanan}, {Nash}, {Neilsen},
  {Newman}, {Nitta}, {Odenkirchen}, {Okada}, {Okamura}, {Ostriker}, {Owen},
  {Pauls}, {Peoples}, {Peterson}, {Petravick}, {Pope}, {Pordes}, {Postman},
  {Prosapio}, {Quinn}, {Rechenmacher}, {Rivetta}, {Rix}, {Rockosi}, {Rosner},
  {Ruthmansdorfer}, {Sandford}, {Schneider}, {Scranton}, {Sekiguchi}, {Sergey},
  {Sheth}, {Shimasaku}, {Smee}, {Snedden}, {Stebbins}, {Stubbs}, {Szapudi},
  {Szkody}, {Szokoly}, {Tabachnik}, {Tsvetanov}, {Uomoto}, {Vogeley}, {Voges},
  {Waddell}, {Walterbos}, {Wang}, {Watanabe}, {Weinberg}, {White}, {White},
  {Wilhite}, {Wolfe}, {Yasuda}, {York}, {Zehavi}, \& {Zheng}}]{Stoughton2002}
{Stoughton}, C., {et~al.} 2002, \aj, 123, 485

\bibitem[{{Tojeiro} {et~al.}(2007){Tojeiro}, {Heavens}, {Jimenez}, \&
  {Panter}}]{Tojeiro2007}
{Tojeiro}, R., {Heavens}, A.~F., {Jimenez}, R., \& {Panter}, B. 2007, \mnras,
  381, 1252

\bibitem[{{Tojeiro} {et~al.}(2009){Tojeiro}, {Wilkins}, {Heavens}, {Panter}, \&
  {Jimenez}}]{Tojeiro2009}
{Tojeiro}, R., {Wilkins}, S., {Heavens}, A.~F., {Panter}, B., \& {Jimenez}, R.
  2009, \apjs, 185, 1

\bibitem[{{Tremonti} {et~al.}(2004){Tremonti}, {Heckman}, {Kauffmann},
  {Brinchmann}, {Charlot}, {White}, {Seibert}, {Peng}, {Schlegel}, {Uomoto},
  {Fukugita}, \& {Brinkmann}}]{Tremonti2004}
{Tremonti}, C.~A., {et~al.} 2004, \apj, 613, 898

\bibitem[{{Tremonti} {et~al.}(2007){Tremonti}, {Moustakas}, \&
  {Diamond-Stanic}}]{Tremonti2007}
{Tremonti}, C.~A., {Moustakas}, J., \& {Diamond-Stanic}, A.~M. 2007, \apjl,
  663, L77

\bibitem[{{Tumlinson} {et~al.}(2011){Tumlinson}, {Thom}, {Werk}, {Prochaska},
  {Tripp}, {Weinberg}, {Peeples}, {O'Meara}, {Oppenheimer}, {Meiring}, {Katz},
  {Dav{\'e}}, {Ford}, \& {Sembach}}]{Tumlinson2011}
{Tumlinson}, J., {et~al.} 2011, Science, 334, 948

\bibitem[{{U} {et~al.}(2009){U}, {Urbaneja}, {Kudritzki}, {Jacobs}, {Bresolin},
  \& {Przybilla}}]{U2009}
{U}, V., {Urbaneja}, M.~A., {Kudritzki}, R.-P., {Jacobs}, B.~A., {Bresolin},
  F., \& {Przybilla}, N. 2009, \apj, 704, 1120

\bibitem[{{Urbaneja} {et~al.}(2005){Urbaneja}, {Herrero}, {Kudritzki},
  {Najarro}, {Smartt}, {Puls}, {Lennon}, \& {Corral}}]{Urbaneja2005}
{Urbaneja}, M.~A., {Herrero}, A., {Kudritzki}, R.-P., {Najarro}, F., {Smartt},
  S.~J., {Puls}, J., {Lennon}, D.~J., \& {Corral}, L.~J. 2005, \apj, 635, 311

\bibitem[{{Weiner} {et~al.}(2009){Weiner}, {Coil}, {Prochaska}, {Newman},
  {Cooper}, {Bundy}, {Conselice}, {Dutton}, {Faber}, {Koo}, {Lotz}, {Rieke}, \&
  {Rubin}}]{Weiner2009}
{Weiner}, B.~J., {et~al.} 2009, \apj, 692, 187

\bibitem[{{Werk} {et~al.}(2011){Werk}, {Putman}, {Meurer}, \&
  {Santiago-Figueroa}}]{Werk2011}
{Werk}, J.~K., {Putman}, M.~E., {Meurer}, G.~R., \& {Santiago-Figueroa}, N.
  2011, \apj, 735, 71

\bibitem[{{Werk} {et~al.}(2010){Werk}, {Putman}, {Meurer}, {Thilker}, {Allen},
  {Bland-Hawthorn}, {Kravtsov}, \& {Freeman}}]{Werk2010}
{Werk}, J.~K., {Putman}, M.~E., {Meurer}, G.~R., {Thilker}, D.~A., {Allen},
  R.~J., {Bland-Hawthorn}, J., {Kravtsov}, A., \& {Freeman}, K. 2010, \apj,
  715, 656

\bibitem[{{Whitaker} {et~al.}(2012){Whitaker}, {van Dokkum}, {Brammer}, \&
  {Franx}}]{Whitaker2012}
{Whitaker}, K.~E., {van Dokkum}, P.~G., {Brammer}, G., \& {Franx}, M. 2012,
  \apjl, 754, L29

\bibitem[{{Wijesinghe} {et~al.}(2012){Wijesinghe}, {Hopkins}, {Brough},
  {Taylor}, {Norberg}, {Bauer}, {Brown}, {Cameron}, {Conselice}, {Croom},
  {Driver}, {Grootes}, {Jones}, {Kelvin}, {Loveday}, {Pimbblet}, {Popescu},
  {Prescott}, {Sharp}, {Baldry}, {Sadler}, {Liske}, {Robotham}, {Bamford},
  {Bland-Hawthorn}, {Gunawardhana}, {Meyer}, {Parkinson}, {Drinkwater},
  {Peacock}, \& {Tuffs}}]{Wijesinghe2012}
{Wijesinghe}, D.~B., {et~al.} 2012, \mnras, 423, 3679

\bibitem[{{Wise} {et~al.}(2012){Wise}, {Abel}, {Turk}, {Norman}, \&
  {Smith}}]{Wise2012}
{Wise}, J.~H., {Abel}, T., {Turk}, M.~J., {Norman}, M.~L., \& {Smith}, B.~D.
  2012, \mnras, 427, 311

\bibitem[{{Worthey} {et~al.}(1994){Worthey}, {Faber}, {Gonzalez}, \&
  {Burstein}}]{Worthey1994}
{Worthey}, G., {Faber}, S.~M., {Gonzalez}, J.~J., \& {Burstein}, D. 1994,
  \apjs, 94, 687

\bibitem[{{Xiao} {et~al.}(2012){Xiao}, {Wang}, {Wang}, {Zhou}, {Lu}, \&
  {Dong}}]{Xiao2012}
{Xiao}, T., {Wang}, T., {Wang}, H., {Zhou}, H., {Lu}, H., \& {Dong}, X. 2012,
  \mnras, 421, 486

\bibitem[{{Yates} {et~al.}(2012){Yates}, {Kauffmann}, \& {Guo}}]{Yates2012}
{Yates}, R.~M., {Kauffmann}, G., \& {Guo}, Q. 2012, \mnras, 422, 215

\bibitem[{{Yoshida} {et~al.}(2011){Yoshida}, {Kawabata}, \&
  {Ohyama}}]{Yoshida2011}
{Yoshida}, M., {Kawabata}, K.~S., \& {Ohyama}, Y. 2011, \pasj, 63, 493

\bibitem[{{Zahid} {et~al.}(2012{\natexlab{a}}){Zahid}, {Bresolin}, {Kewley},
  {Coil}, \& {Dav{\'e}}}]{Zahid2012a}
{Zahid}, H.~J., {Bresolin}, F., {Kewley}, L.~J., {Coil}, A.~L., \& {Dav{\'e}},
  R. 2012{\natexlab{a}}, \apj, 750, 120

\bibitem[{{Zahid} {et~al.}(2012{\natexlab{b}}){Zahid}, {Dima}, {Kewley}, {Erb},
  \& {Dav{\'e}}}]{Zahid2012b}
{Zahid}, H.~J., {Dima}, G.~I., {Kewley}, L.~J., {Erb}, D.~K., \& {Dav{\'e}}, R.
  2012{\natexlab{b}}, \apj, 757, 54

\bibitem[{{Zahid} {et~al.}(2013{\natexlab{a}}){Zahid}, {Geller}, {Kewley},
  {Hwang}, {Fabricant}, \& {Kurtz}}]{Zahid2013b}
{Zahid}, H.~J., {Geller}, M.~J., {Kewley}, L.~J., {Hwang}, H.~S., {Fabricant},
  D.~G., \& {Kurtz}, M.~J. 2013{\natexlab{a}}, \apjl, 771, L19

\bibitem[{{Zahid} {et~al.}(2011){Zahid}, {Kewley}, \& {Bresolin}}]{Zahid2011a}
{Zahid}, H.~J., {Kewley}, L.~J., \& {Bresolin}, F. 2011, \apj, 730, 137

\bibitem[{{Zahid} {et~al.}(2013{\natexlab{b}}){Zahid}, {Yates}, {Kewley}, \&
  {Kudritzki}}]{Zahid2013a}
{Zahid}, H.~J., {Yates}, R.~M., {Kewley}, L.~J., \& {Kudritzki}, R.~P.
  2013{\natexlab{b}}, \apj, 763, 92

\bibitem[{{Zaritsky}(1994)}]{Zaritsky1994b}
{Zaritsky}, D. 1994, \aj, 108, 1619

\bibitem[{{Zhang} \& {Thompson}(2010)}]{Zhang2010}
{Zhang}, D., \& {Thompson}, T.~A. 2010, ArXiv e-prints

\bibitem[{{Zu} {et~al.}(2011){Zu}, {Weinberg}, {Dav{\'e}}, {Fardal}, {Katz},
  {Kere{\v s}}, \& {Oppenheimer}}]{Zu2011}
{Zu}, Y., {Weinberg}, D.~H., {Dav{\'e}}, R., {Fardal}, M., {Katz}, N., {Kere{\v
  s}}, D., \& {Oppenheimer}, B.~D. 2011, \mnras, 412, 1059

\bibitem[{{Zurita} \& {Bresolin}(2012)}]{Zurita2012}
{Zurita}, A., \& {Bresolin}, F. 2012, \mnras, 427, 1463

\end{thebibliography}

 \end{document}